\documentclass[%
groupedaddress,
 amsmath,amssymb,
twocolumn,floatfix,aps,
]{revtex4-1}

\usepackage{graphicx}
\usepackage{dcolumn}
\usepackage{bm}
\usepackage{xcolor}
\usepackage{float}
\usepackage[caption = false]{subfig}
\usepackage[pdftex]{hyperref}
\usepackage{bbm}

\DeclareMathOperator{\sech}{sech}

\def\be{\begin{equation}}
\def\ee{\end{equation}}

\def\bp{\begin{pmatrix}}
\def\ep{\end{pmatrix}}

\begin{document}

\title{Interplay of Solitons and Radiation in One-Dimensional Bose Gases}

\author{Yuan Miao}
\author{Enej Ilievski}
\author{Oleksandr Gamayun} 
 
\affiliation{
 Institute for Theoretical Physics, Institute of Physics and Delta Institute for Theoretical Physics, Universiteit van Amsterdam, Science Park 904, 1098XH Amsterdam, the Netherlands
}

\date{\today}

\begin{abstract}
We study relaxation dynamics in one-dimensional Bose gases, formulated as an initial value problem for the classical
nonlinear Schr\"{o}dinger equation. We propose an analytic technique which takes into account the exact spectrum of nonlinear modes, that is both soliton excitations and dispersive continuum of radiation modes.
Our method relies on the exact large-time asymptotics and uses the so-called dressing transformation to account for the solitons. The obtained results are quantitatively compared with the predictions of the linearized approach in the framework of the
Bogoliubov theory. In the attractive regime, the interplay between solitons and radiation yields a damped oscillatory motion of the profile which resembles breathing. For the repulsive interaction, the solitons are confined in the sound cone region separated from the supersonic radiation.
\end{abstract}

\maketitle


\section{Introduction}
\label{sec:intro}

Recent years have brought tremendous experimental and theoretical progresses in understanding various aspects of equilibrium and nonequilibrium physics in strongly correlated many-body systems, especially in the domain of quantum gases~\cite{bloch2008many, Jaksch_2005, kinoshita2006quantum, Hofferberth_2007, Langen_2015}. There has been a particularly intense focus on studying relaxation phenomena and microscopic mechanisms responsible for thermalization in isolated many-body systems~\cite{Deutsch_1991, Srednicki_1994, Rigol_2008, Rigol_2007, Caux_2013, Eisert_2015, Vidmar_2016, Gogolin_2016, D_Alessio_2016, Caux_2016, Cugliandolo_2018, Deutsch_2018}. Equilibration in generic chaotic (i.e. ergodic) systems is nowadays quite well understood, primarily using
the arguments of Eigenstate Thermalization Hypothesis~\cite{Deutsch_1991, Srednicki_1994, Rigol_2008, Alba_2015, D_Alessio_2016, Deutsch_2018}. On the other hand, it has been argued that nonergodic systems, e.g. models which lie in the proximity of an integrable point, fail to thermalize in the conventional sense owing to an extensive amount of local conservation law which severely constraints the dynamics.
The proposed generalized Gibbs ensembles~\cite{Rigol_2007, Vidmar_2016,EF16_review} have subsequently been scrutinized in a variety of noninteracting and interacting exactly solvable quantum many-body dynamics~\cite{Caux_2012,Ilievski_2015, Essler_2015, Ilievski_2016, Ilievski_2017}, and in the context of classical integrable systems by taking the classical limit of quantum fields~\cite{De_Luca_2016}.

Theoretical studies of nonequilibrium phenomena in exactly solvable models are for the most part concerned with steady states and their properties, while much less is known about the relaxation dynamics at large (or intermediate) time-scales. Despite integrability, the latter represents a formidable task in both classical and quantum many-body systems with interacting degrees of freedom.

For instance, while the theoretical framework for solving integrable differential 
equations describing classical field theories is very well developed, a full-fledged analytic treatment of nonlinear wave equations 
is unfortunately not tractable in full generality, explaining the scarcity of closed-form results in the literature.
This work aims to partially fill this gap by presenting some nontrivial analytic results in a physically relevant setting.
In particular, we consider interacting one-dimensional~(1D) Bose gases which are, in the weakly-coupled regime, well described
by the classical nonlinear Schr\"{o}dinger equation (NLSE) (also known as the Gross--Pitaevskii equation),
one of the prime examples of exactly solvable nonlinear wave equations~\cite{Pitaevskii_2016, Pethick_2001, Faddeev, novikov1984theory}.

We study the initial value problem for the NLSE by implementing a `classical quench protocol', i.e. initializing
an inhomogeneous  profile and letting it evolve under the nonlinear evolution law. By exploiting integrability of
the equation of motion, we devise an analytic technique by combining the (Darboux) dressing transformation and exact asymptotic formulae, which allows us to  accurately approximate the evolution even on moderately short time scales. Our results are benchmarked against the standard linear approximation and numerical simulations.

We consider field configurations which decay towards a constant vacuum density at large distance. The setup we study therefore differs from the conventional setting in quantum quenches which address local equilibration in thermodynamic quantum gases at finite density. Nonetheless, it turns out that an extensive amount of conservation laws once again play a pivotal role in constraining the relaxation process.

Integrable classical field theories in general feature two types of solutions with a distinct character: (i) nonlinear interacting particles known as solitons, representing nondispersive localized field configurations, and (ii) a continuum of nonlinear dispersive modes called radiation. It is common practice however to treat small fluctuations of a uniform background density within the linear theory of noninteracting (Bogoliubov) quasi-particles. Contrary to the solitons and radiation modes the latter are not the proper elementary modes of the NLSE, and it is thus natural to wonder whether they remain a meaningful concept in genuine far-from-equilibrium scenarios such as classical quenches considered in this paper. In an attempt to answer this question, we investigate the linearized dynamics at a qualitative and quantitative level, focusing on intermediate time-scales where the effects of nonlinearity cannot be neglected.

Aside from the theoretical interest, it is worthwhile to briefly mention some experimental aspects of the quench protocol proposed here. For nearly two decades, the experimental realizations of Bose-Einstein condensation (BEC)~\cite{anderson1995observation, davis1995bose, bradley1995evidence} offer new routes for investigating the many-body phenomena in a controllable and precise manner. In particular, 1D Bose gases with tunable s-wave scattering interaction (induced by Feshbach resonance)~\cite{burger1999dark, PhysRevLett.89.110401, khaykovich2002formation, lepoutre2016production} are of great interest, due to the simple and exactly solvable theoretical models which displaying rich physical behavior. Theoretically speaking, 1D Bose gases with s-wave scattering can be modelled as the Lieb-Liniger model~\cite{Lieb_Liniger_1963, Lieb_1963}, which is, in the weak-coupling (Gross--Pitaevskii) limit, nothing but a 1D NLSE~\cite{Pethick_2001,Pitaevskii_2016,carr2000stationary1, carr2000stationary2}. Above all, solitonic excitations in both 1D attractive and repulsive NLSE have been observed experimentally~\cite{burger1999dark, khaykovich2002formation, Becker, PhysRevLett.101.130401, Stellmer2008, lepoutre2016production}, making it promising to realize our quench proposal in ultracold atom experiments.

The paper is organized as follows. In Sec.~\ref{sec:brightS} we consider quenches in the attractive 1D NLSE.
In Sec.~\ref{subsec:solitonless_bright} we derive the long-time asymptotic solutions for a solitonless quench, while in
Sec.~\ref{subsec:BdG_bright} we describe quenches when both the soliton and radiation modes are present, and discuss
the ``soliton breathing'' effect.

The quench in the repulsive NLSE with finite density is studied in Sec.~\ref{sec:dark}, revealing different physical phenomena
such as e.g. the separation of the soliton sound cone from the supersonic radiation modes.
We conclude in Sec.~\ref{sec:conc} by summarizing the main results and listing some open questions. A concise introduction to
the classical inverse scattering method and a detailed derivation of the complete spectrum of the linearized
equations are presented in the appendices.

\section{Attractive Interaction}
\label{sec:brightS}

We consider an attractive (focusing) NLSE, which in the dimensionless unit ($\hbar = 2m = 1$) takes the form
\be 
i \partial_t \psi(x,t) = -  \partial_x^2 \psi(x,t) + 2 \varkappa |\psi(x,t)|^2 \psi(x,t) .
\label{NLSE_attractive}
\ee

This equation can be solved by the inverse scattering method, which we briefly describe below using Refs.~\cite{Faddeev, ZMNP}. The essential idea is to reformulate the original nonlinear equation as an auxiliary linear problem for the `wave-function' $F=(F_{1},F_{2})^{\rm T}$,
\be\label{ALP}  
\frac{\mathrm{d}}{\mathrm{d} x}\left(\begin{array}{c}
	F_1 \\
	F_2
\end{array}\right)  = \left[\frac{\lambda \sigma_3}{2i}+U_{\psi} \right]\left(\begin{array}{c}
F_1 \\
F_2
\end{array}\right),
\ee
with 
\be \label{upsi}
\sigma_3 = \left(\begin{array}{cc}
	1& 0\\
	0 & -1
\end{array}\right),\qquad
U_{\psi} =\sqrt{|\varkappa |} \left(\begin{array}{cc}
0 & i\bar{\psi} \\
i\psi & 0
\end{array}\right),
\ee
which is interpreted as a scattering problem: the field $\psi(x)$ is viewed as the scattering potential,
while the spectral parameter $\lambda$ plays the role of energy. The transfer matrix for this problem can be presented as 
\be 
T(\lambda) = \left(\begin{array}{cc}
	a(\lambda) & -\bar{b}(\lambda) \\
	b(\lambda) & \bar{a}(\lambda)
\end{array}\right),
\ee
with ${\rm det} \, T(\lambda) = |a(\lambda)|^{2}+|b(\lambda)|^{2}=1$. 
One can compute the time dependence of these scattering data assuming that $\psi(x,t)$ satisfies NLSE \eqref{NLSE_attractive},
obtaining a remarkably simple dependence \cite{Faddeev}:
\be\label{evolT}
a(\lambda,t) = a(\lambda,0),\qquad b(\lambda,t) = e^{-i \lambda^2 t}b(\lambda,0).
\ee 
This way, to solve the initial value (Cauchy) problem of Eq. \eqref{NLSE_attractive}, one has to compute and diagonalize
the transfer matrix $T(\lambda)$ for a given initial profile, evolve the transfer matrix according to Eq.~\eqref{evolT} and, finally,
retrieve the time-evolved potential from the scattering data.
The last step is referred to as the ``Inverse Scattering" and requires to solve the Gelfand--Levitan--Marchenko
linear integral equation.
In the very special cases of the reflectionless potentials, characterized by $b(\lambda) =0$,
this equation can be solved analytically and its solutions are called {\it solitons}.
In particular, the one-soliton solution reads
\be 
\psi_s (x,t) = \frac{u}{\sqrt{|\varkappa |}} \frac{\exp \left[i(\varphi_0 + vx +(u^2-v^2)t )\right]}{\cosh \left[ u (x-2vt -x_0)  \right]},
\label{bright_soliton}
\ee
describing a bell-shaped profile traveling with the velocity $2v$, usually named as a bright soliton.  Without loss of generality, 
one can set the initial positions $\varphi_0 = 0$, $x_0 = 0$, the inverse width $u=1$ as well as the velocity $2v=0$ by an appropriate Galilean transformation. For generic initial conditions, i.e. $b(\lambda)\neq 0$, in addition to solitons the spectrum also
contains {\it radiation} modes, which unlike solitons are dispersive.

In order to model various types of initial conditions while still being able to perform analytic computations, we consider a rescaled Satsuma-Yajima profile \cite{SatsumaYajima,Miles,gamayun2016soliton} with $\eta \in \mathbb{R}^+$
\be \label{initialAtrractive}
\psi(x,0) = \frac{1}{\sqrt{|\varkappa |} \cosh(x/\eta)} .
\ee
The quench parameter $\eta>0$ can be removed from the profile by a suitable rescaling of Eq.~\eqref{NLSE_attractive} which will change the nonlinearity coefficient (the coupling constant), $\varkappa \to \varkappa^{\prime}=\varkappa/\sqrt{\eta}$. Hence, it is feasible to realize this profile in the ultracold atom experiments by preparing a one-soliton state and subsequently quenching parameters of the holding trap and external fields to induce the change in the coupling constant~\cite{gamayun2015fate,gamayun2016soliton,caudrelier2016quench,franchini2016hydrodynamics}.
One can think of this as a classical analog of the interaction (nonlinearity coefficient) quench in the Lieb-Liniger model~\cite{Kormos_2013, De_Nardis_2014, Piroli_2016}.

The scattering data of potential \eqref{initialAtrractive} can be computed exactly\cite{gamayun2015fate,gamayun2016soliton}
\begin{align}
\label{asol}
a(\lambda) &= \frac{\Gamma(\frac{1}{2}-\frac{i \lambda \eta}{2})^2}{\Gamma(\frac{1}{2}-\frac{i \lambda \eta}{2}-\eta)\Gamma(\frac{1}{2}-\frac{i \lambda \eta}{2}+\eta)},\\
\label{bb} 
b(\lambda) &= \frac{i \sin (\pi \eta)}{\cosh(\pi \eta \lambda/2)}.
\end{align}
For integer quench parameters $\eta = n \in \mathbb{N}$, corresponding to the reflectionless potential $b(\lambda) = 0$,
the solution only involves solitons. The soliton parameters are identified with the zeros of $a(\lambda)$ in the upper half plane,
which has the form
\be 
a(\lambda) = \prod_{k=1}^n \frac{\lambda - (2k-1)i/n}{\lambda + (2k-1)i/n},
\ee
corresponding to an $n$-soliton solution. In this scenario, all solitons have zero velocity and the time-evolution exhibits
a periodically oscillating behavior due to their mutual interaction which we refer to as ``soliton 
breathing''~\cite{gamayun2016soliton}.

In the generic case, we put $m= \lfloor \eta + 1/2 \rfloor $ (the greatest integer smaller or equal to $\eta+1/2$), and the factor $a(\lambda)$ can be presented as
\be \label{a(l)BR}
a(\lambda) = \prod_{k=1}^{m} \frac{\lambda - \lambda_k}{\lambda - \bar{\lambda}_k} \exp\left[
\int\limits_{-\infty}^\infty \frac{\log(1-|b(\mu)|^2)}{\mu-\lambda-i0}  \frac{\mathrm{d} \mu}{2\pi i} \right],
\ee
where
\be 
\lambda_k = i \frac{2 \eta - 2 k +1}{\eta}. 
\ee
This representation indicates that the field profile involves $m$ solitons (described by parameters $\lambda_k$), superimposed on a continuous background of radiation modes encoded by the nonzero reflection coefficient $r(\lambda)\equiv b(\lambda)/a(\lambda)$.

We focus our analysis on two representative regimes: (A) $0<\eta < 1/2$; (B)~$1/2<\eta < 3/2$. In the (A) case, solitons are absent and the evolution is governed solely by radiation modes, resulting in a ballistic widening of the initial profile similar to the wave-packet spreading in the linear Schr\"odinger equation. In the (B) case, the nonlinearity becomes much more important since the profile contains a soliton. Therefore, in addition to the ballistic expansion there remains a ``stable'' part. The attractive 
interaction between the soliton and radiation will show up in the form of ``breathing'' of the whole profile, similarly to the two-soliton solution \cite{gamayun2016soliton}. In the next section we examine these two regimes using an analytic expression for the time-asymptotic behavior and give a quantitative analysis of the soliton breathing phenomenon. Furthermore, we compare our analytic findings with numerical integration.

\begin{figure}
	\centering
	\includegraphics[width=\linewidth]{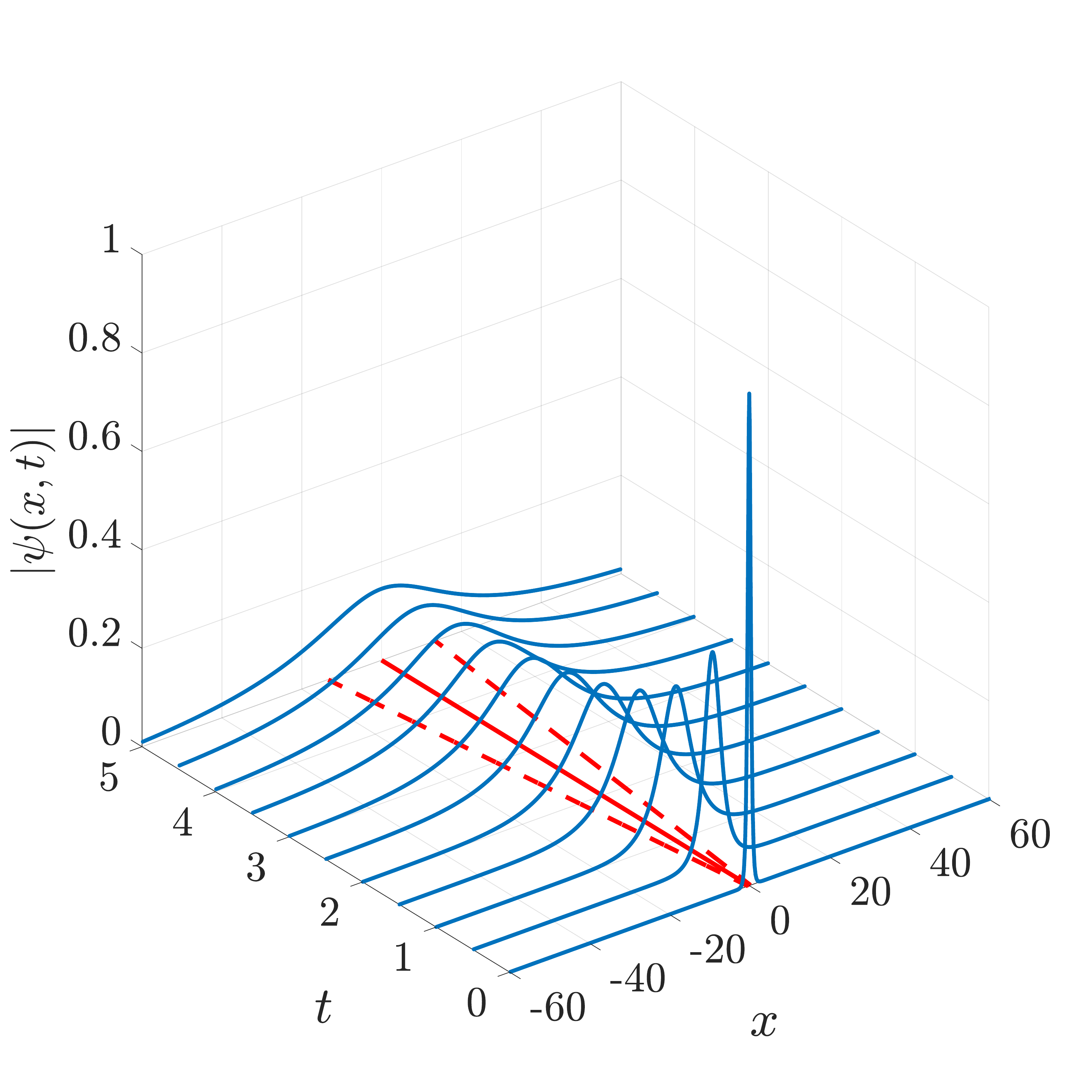}
	\caption{Time evolution of quenched profile~\eqref{initialAtrractive} with $\varkappa = -1$ and $\eta = 1/3$. The dashed lines denote the half width of the profile, indicating a ballistic (i.e. linear in time) expansion of the half width.}
	\label{fig:ballistic}
\end{figure}	

\subsection{\label{subsec:solitonless_bright}Solitonless quench} 

A typical time evolution of a quenched profile~\eqref{initialAtrractive} with $\eta <1/2$ is plotted in Fig.~\ref{fig:ballistic}.
One can observe a ballistic expansion, reminiscent of the wave-packet spreading governed by the linear Schr\"odinger equation.

Indeed, at the phenomenological level the situation is described as follows. At the initial moment of time a very narrow profile can be rescaled to the unit characteristic width by the rescaling $x\to \eta\,x$ and $t \to \eta^2\,t$, resulting in the nonlinearity 
coefficient proportional to $\eta^2$ and the dynamics dominated by the linear part of the equation. 
The evolution causes the spreading amplitude damping of the profile, thus suppressing the importance of the nonlinear terms.
The complete account of the nonlinear terms needed for the analysis of the exact asymptotic expression requires sophisticated 
techniques based on the Riemann--Hilbert problem \cite{zakharov1976asymptotic,DeiftP.A.1993}. 
We shall instead proceed with a more physically transparent (but less rigorous) method, inspired by the above analogy with the linear case.

We begin by recalling the general solution of the linear Schr\"odinger equation and its large-time asymptotics,
\be \label{psi2}
\psi(x,t) = \int \frac{\mathrm{d} k}{\sqrt{2\pi}}   f(k)e^{ik x-ik^2t}  \approx \frac{f(x/2t)}{\sqrt{2it}}e^{ix^2/4t}.
\ee
Function $f(k)$ is uniquely determined from the initial condition. This result motivates to look for the solution of Eq. \eqref{NLSE_attractive} in the following form \cite{ablowitz1976asymptotic1, segur1976asymptotic2,zakharov1976asymptotic}
\be \label{asymptBR}
\begin{split}
	\psi(x,t) & = \frac{1}{\sqrt{t}} \left[ f+\sum^{\infty}_{n=1} \sum^{n}_{k=0} \frac{(\log t)^k}{t^n} f_{nk}\right] \\ 
	& \times\exp \left(\frac{i x^2}{4t}+i \Xi \log t\right),
\end{split}
\ee
where $f$, $f_{nk}$ and $\Xi$ are all functions of the scaling variable $x/2 t$.
Substituting this expression into Eq.~\eqref{NLSE_attractive}, we find that all functions can be expressed in terms of the amplitude function $f(x/2t)$ and its derivative, and in particular
\be 
\Xi(x/2t) = -2\varkappa |f(x/2t)|^2.
\ee

The aim now is to determine $f(x/2 t)$. In order to relate it to the initial profile, we borrow the logic
of Ref.~\cite{ablowitz1976asymptotic1}, and compute the local conserved charges both on the initial profile and on the asymptotic expression \eqref{asymptBR}.

According to Eq. \eqref{evolT}, $\log a(\lambda)$ is a conserved quantity and the conventional integrals of motions
are defined as coefficients in the asymptotic expansion, namely
\be \label{consvA}
\log a(\lambda)  = i \varkappa \sum\limits_{k=1}^\infty \frac{Q_n}{\lambda^n} + \mathcal{O}(|\lambda|^{-\infty}).
\ee 
Taking into account Eq. \eqref{a(l)BR}, the conserved charges can be found from the moment expansion
\be\label{QN1}
Q_n = \int \frac{\log(1-|b(\mu)|^2)}{2\pi \varkappa} \mu^{n-1} \mathrm{d} \mu.
\ee
On the other hand, the local conserved charges can be presented as spatial integrals of local densities
\be 
Q_n = \int \mathrm{d} x \rho_n(x),
\label{Qdens}
\ee
which could be found from the recurrence relation \cite{Faddeev,doi:10.1137/1.9781611970883}
\be 
\rho_{n+1} = - i \bar{\psi} \partial_x \frac{\rho_n}{\bar{\psi}} + \varkappa \sum\limits_{k=1}^{n-1}\rho_{n-k}\rho_k,\qquad
\rho_1 = |\psi |^2.
\ee

\begin{figure}
	\centering
	\includegraphics[width=\linewidth]{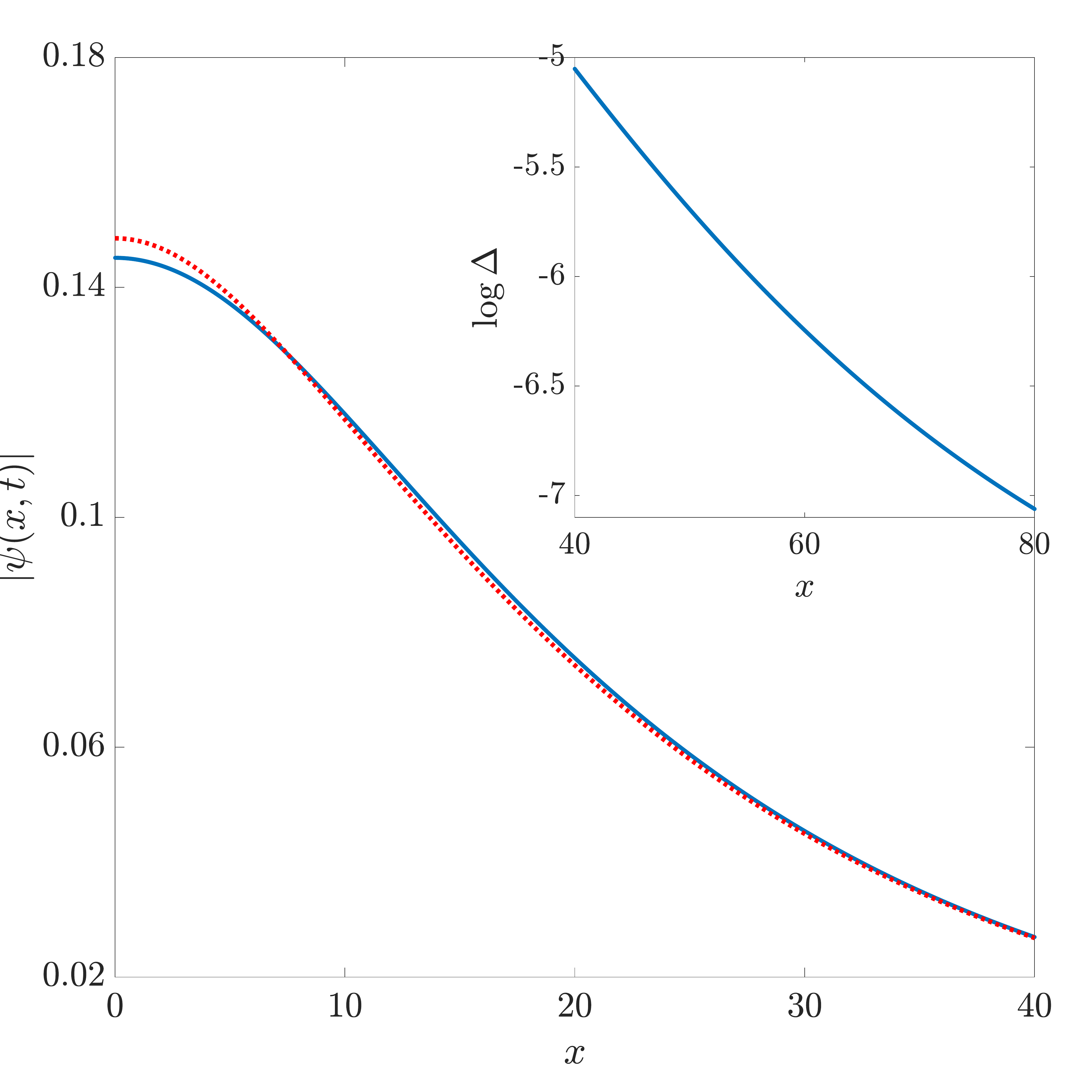}
	\caption{Time evolved profile \eqref{initialAtrractive} for $\varkappa = -1$ and $\eta=1/3$ at $t=5$, obtained by numerical integration (solid line) 
	and from the asymptotic solution $\psi_a$ given by Eq.~\eqref{bright_asymp_full} (dashed line). The inset plot shows the 	
	relative difference between the two integration methods $\Delta = |\psi(x,t)-\psi_a(x,t)|/|\psi(x,t)|$.}
	\label{fig:diff_1_over_3}
	\label{BrightASYMP1}
\end{figure}

Since asymptotically the nonlinear part is suppressed in the inverse powers of $t$, the conserved densities will be
approximately the same as in the linear case. Evaluating them on the profile \eqref{asymptBR}, one gets
\be 
\rho_{n} \approx \bar{\psi}(-i\partial_x)^{n-1} \psi =  \left(\frac{x}{2 t} \right)^{n-1} \frac{|f|^2}{t}.
\ee 
After the change of variable, $\mu = x/2t$, the conserved charges read
\be 
Q_{n} = 2 \int \mu^{n-1} |f(\mu)|^2 \mathrm{d} \mu + o (1).
\ee
Hence, taking into account Eq. \eqref{QN1}, we readily identify 
\be 
|f( \mu)|^2 = \frac{\log(1-|b(\mu)|^2)}{4 \pi \varkappa} , \label{fa}
\ee
which in turn allows to completely determine the asymptotic solution from the extensive number of the conserved charges~(i.e. the \textit{action variables} for the 1D NLSE).

The determination of the phase of $f(\mu)$ is more technical, and the result reads~\cite{zakharov1976asymptotic}

\begin{align} 
	{\rm arg}\,f( \mu) &=  \frac{3\pi}{4}-{\rm arg}\,b(\mu) - \mathrm{arg} \, \Gamma (4 \pi i |f(\mu)|^2 ) \nonumber \\
	&- 2 \varkappa |f(\mu)|^2 \log 2 \nonumber \\
	&+ 4 \int_{-\infty}^{\mu } \log \left( \mu - \mu^{\prime} \right) \mathrm{d} (|f(\mu^{\prime})|^2) \nonumber \\
	&- 4 \int^{\infty}_{\mu} \log \left( \mu^{\prime} - \mu \right) \mathrm{d} (|f(\mu^{\prime})|^2) + o(1) .
	\label{o1phase}
\end{align}

The phase provides the \textit{angle variables} for the 1D NLSE~\cite{A.R.Its1988}. The full expression for the asymptotic
solution reads
\be 
	\psi_a (x,t) = \frac{1}{\sqrt{t}} \left| f\left( \frac{x}{2 t} \right) \right| \exp \left[ i \theta_a \left( \frac{x}{2 t} , t \right) \right] ,
	\label{bright_asymp_full}
\ee
with the phase
\be 
	\theta_a \left( \frac{x}{2 t} , t \right) = \frac{x^2}{4 t} + \Xi \left( \frac{x}{2 t} \right) \log t + \arg f \left( \frac{x}{2 t} \right) .
	\label{bright_asymp_phase}
\ee

Noticing that the density of the asymptotic profile
is fully characterized by the modulus $|f(\mu)|^{2}$, i.e. local properties (charges) of individual initial configurations, our scenario closely resembles the problem of identifying a generalized Gibbs ensemble which corresponds to the reduced density matrix in the steady-state limit of a quantum quench.
The information about the asymptotic phase \eqref{bright_asymp_phase} cannot be fully restored from the integrals of motion but requires extra information about the initial angle coordinates.
Nevertheless, by discarding the information about the initial phase (e.g. 
by uniformly averaging over it), one can define a microcanonical ensemble of states whose long-time asymptotics is completely determined by the initial values of the conserved charges.
In other words, the (microcanonical) ergodic average of local observables (e.g. observables proportional to field $\psi(x,t)$ and its derivatives) only retains information about the initial action variables, and this can be extracted via the method presented here.

In Fig. \ref{BrightASYMP1} we compare the asymptotic expression with the results of numerical integration. It is worth stressing that the asymptotic solution~\eqref{bright_asymp_full} to Eq.~\eqref{NLSE_attractive} only becomes exact when all higher order terms in Eq.~\eqref{asymptBR} are taken into account. Remarkably, we observe no dramatic effect of nonlinearity coefficient $\varkappa$ on the half-width of the quenched profile, which means that the linearized solution offers reasonably good results for all times if the initial profile is small enough (of course, the initial profile is proportional to $1/\sqrt{\varkappa}$, but the half-width grows strictly ballistically as in the linear equation). Still, the nonlinearity has to be accounted for in order to give the correct values of the (infinitely many) conserved charges (cf. $\varkappa$ dependence in Eq.~\eqref{fa}).
For a generic $\eta <1/2$ our asymptotic profile is in good qualitative agreement with the exact profile even around the origin $x\approx 0$. 
However, at the soliton-birth threshold $\eta=1/2$ this description manifestly breaks down \cite{Malomed_1987}. Indeed, when $\lambda=x/2t\sim 0$, we have a logarithmic divergence of the 
profile \eqref{bright_asymp_full} since  $\log(1 -|b(\lambda)|^2) \sim \log \lambda^{-2}$. In Ref. \cite{Malomed_1987}, the asymptotic behavior in this region is described by a phenomenological substitution $\lambda^{2}\to \lambda^{2} + \frac{1}{4t}$. The correct asymptotics can be deduced from the corresponding Riemann-Hilbert problem, which requires more technical and elaborate analysis. 

The presented method for the analytic determination of the asymptotic solution can be easily generalized to the other classical quench protocols, as long as the class of initial conditions permits to extract the scattering data, which enables
computation of the conserved charges. We have demonstrated that the asymptotic solution~\eqref{bright_asymp_full} accurately describes the situation of the solitonless initial profiles.
In the presence of solitons, the outlined method cannot be straightforwardly generalized despite the fact
that the phase-space expressions for the local conserved charges remain {\rm exactly} the same  as in solitonless case (see Eq. \eqref{Qdens}). However, from the asymptotic expansion of $\log a(\lambda)$ Eq. \eqref{consvA} using general presentation \eqref{a(l)BR} one can clearly see that the conserved charges acquire soliton corrections.
Therefore, instead of \eqref{asymptBR}, a new ansatz for the asymptotic field profile is needed. The latter can be, in principle, deduced from the large-time asymptotic solution of the NLSE linearized on the soliton background. But even in this case, it is not clear how to effectively evaluate the densities and compute the corresponding charges. Similar approaches developed in
Ref. \cite{segur1976asymptotic2} give unsatisfactory results. Thus, to describe the (B) regime, $\eta > \frac{1}{2}$, we pursue
a different strategy presented in the next section. Firstly, we compute the full time dependence of the NLSE linearized on the soliton background, which yields reasonably good results on short and intermediate times. Secondly, we employ the Darboux transformation, which permits us to satisfy condition \eqref{QQ1} exactly, but requires explicit knowledge of the asymptotics for the radiative part.

\subsection{Interplay of solitons and radiation}
\label{subsec:BdG_bright}

\begin{figure}
	\centering
	\includegraphics[width=\linewidth]{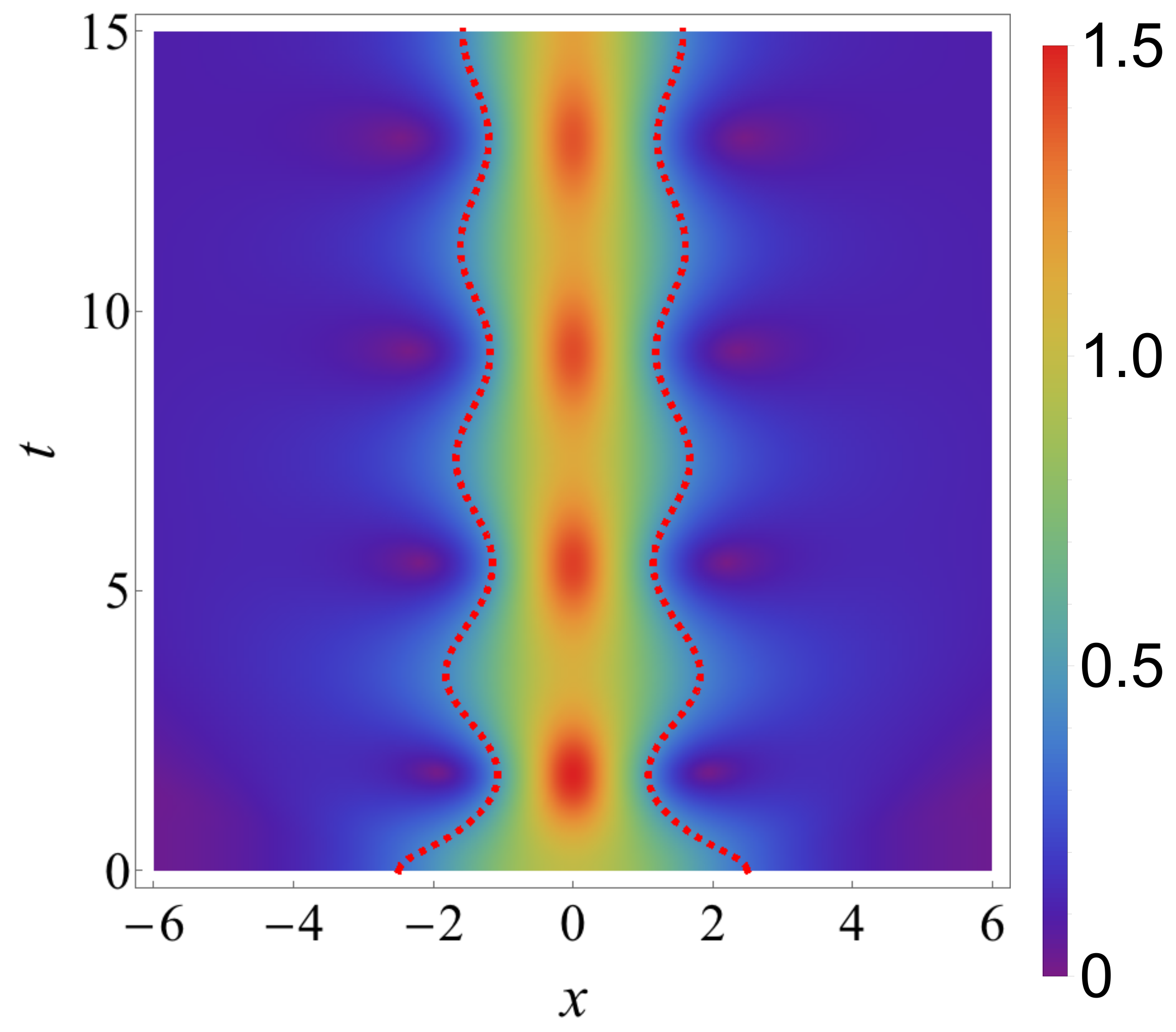}
	\caption{Time evolution of the intensity $|\psi(x,t)|$ of the quenched profile~\eqref{initialAtrractive} with $\varkappa=-1$ and $\eta = \sqrt{2}$, showing persistent oscillations referred to as the ``soliton breathing''. The red lines denote the half width of the profile.}
	\label{fig:bright_breathing}
\end{figure}

\begin{figure}
	\centering
	\includegraphics[width=\linewidth]{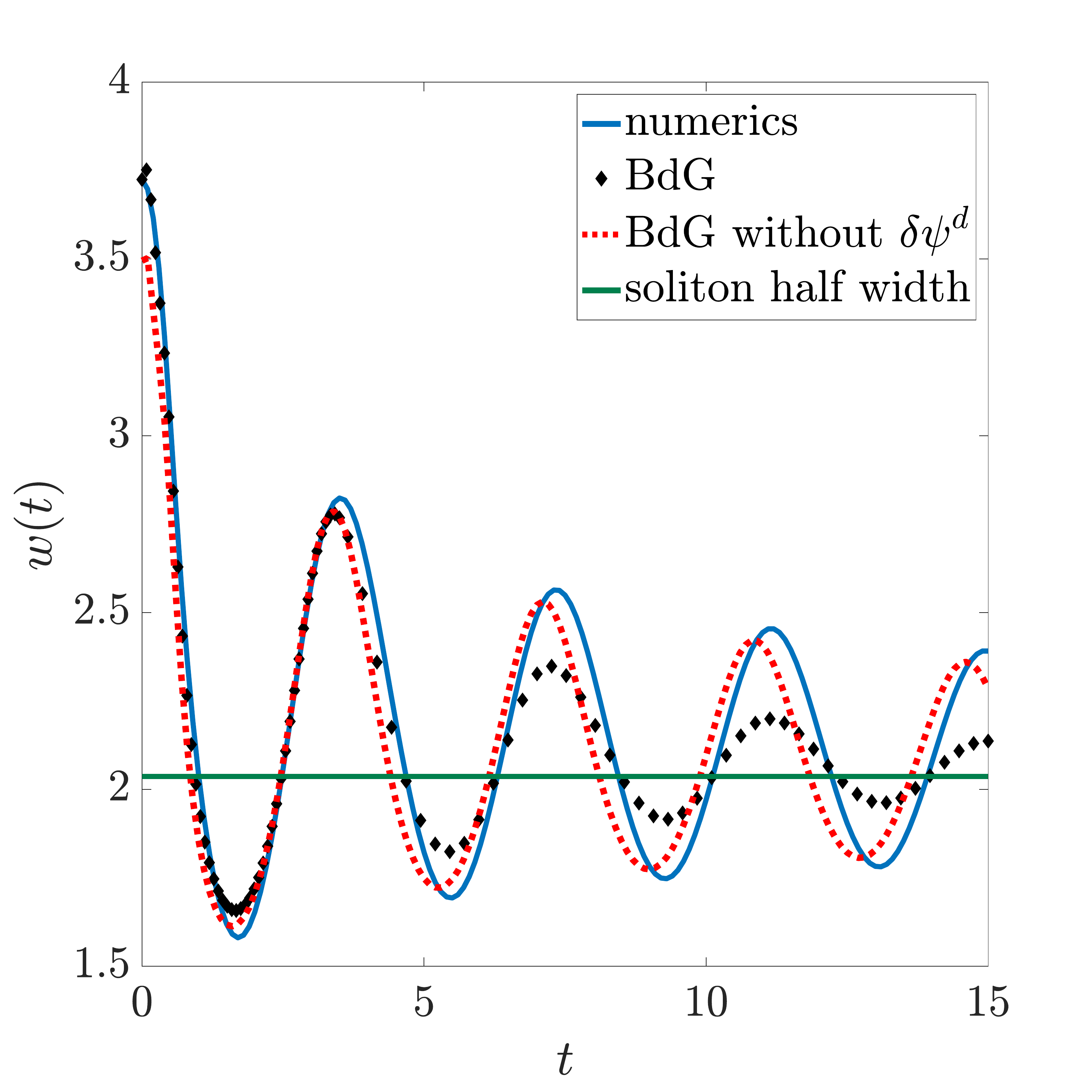}
	\caption{Time dependence of the profile's half-width $w(t)$ obtained from the linearized (BdG) equation with parameter $\varkappa = -1$ and $\eta = \sqrt{2}$, with (diamonds) and without (dashed line) the zero-mode components, compared to the exact numerical integration.}
	\label{fig:half_width_comp}
\end{figure}

The time evolution of the quenched profile undergoes a qualitative change in the regime $\eta>1/2$ where the nonlinear effects become 
more pronounced. For $1/2<\eta<3/2$, according to Eq. \eqref{a(l)BR}, the spectral data contains a single static soliton whose 
form can be explicitly computed from the scattering data~\cite{gamayun2015fate, gamayun2016soliton}
\be\label{psiS} 
\psi_s (x,t) = \frac{\nu}{\sqrt{|\varkappa|}} \frac{\exp (i \nu^2 t)}{\cosh (\nu x)},\qquad \nu = \frac{2\eta -1}{\eta} .
\ee
The results of numerical integration are shown in Fig.~\ref{fig:bright_breathing}. The effect of the attractive interaction is that ballistic spreading of the radiation modes is now accompanied by a ``breathing motion'', i.e. the peak of the soliton oscillates, which, in particular, can be seen in the time-dependence of the profile's half-width, see Fig. \ref{fig:half_width_comp}. The period of oscillations can be approximated by the period of the soliton,
\be 
T \simeq \frac{2\pi}{\nu^2} = \frac{2\sqrt{2}\pi}{2\sqrt{2}-1},
\ee
given that near the origin ($x=0$) the evolved profile is a sum of the soliton and slowly varying component
consisting of the radiation modes.

\paragraph*{Bogoliubov theory.}
In order to present a heuristic picture of this phenomenon, we consider the linearized theory with respect to the soliton solution $\psi_s$ \eqref{psiS}.  Namely, we split 
\be 
\psi(x,t) = \psi_s(x,t) + \delta \psi(x,t),
\ee
and account only for the linear terms in the equation of motion for $\delta \psi (x,t)$, yielding
\be
i\partial_t \delta \psi  = - \partial_x^2 \delta \psi + 4 \varkappa |\psi_s|^2 \delta \psi + 2 \varkappa \psi_s^2 \delta\bar{\psi}.
\label{bdg}
\ee
This equation, presented in the matrix form, is nothing but
the Bogoliubov--de Gennes (BdG) equation~\cite{Bogoljubov_1958,Pitaevskii_2016,walczak2011exact, Kominis_2006, Kominis_2007}
\be
i \partial_t \bp \delta \psi \\ \delta \bar{\psi}  \ep = \bp A & B \\ -\bar{B} & -A \ep \bp \delta \psi \\  \delta\bar{\psi}  \ep ,
\label{linearized2}
\ee
where $A = -\partial_x^2 + 4 \varkappa |\psi_s|^2 $ and $B = 2 \varkappa \psi_s^2$.
In order to get rid of the time dependence which is due to the ``potential'' $\psi_s(x,t)$ in Eq.~\eqref{linearized2}, we introduce
\be 
	\delta \chi(x,t) = e^{-i \nu^2 t} \delta \psi(x,t) ,
\ee 
so the corresponding BdG equation in the rotating frame reads
\be 
	i\partial_t \bp \delta \chi \\ \delta \bar{\chi} \ep = 
	\bp \tilde{A} &  \tilde{B} \\
	- \tilde{B} & -\tilde{A} \ep  \bp \delta \chi \\ \delta \bar{\chi} \ep  ,
	\label{linearized_chi}
\ee
where now $\tilde{A} = A + \nu^2$ and $\tilde{B} = 2 \varkappa |\psi_s|^2$ are independent of time. 
General solutions of BdG equation~\eqref{linearized_chi} can be presented in the form 
\be 
\delta \chi (x,t) = \delta \chi^c (x,t) + \delta \chi^d (x,t) , 
\ee
where $\delta \chi^c$ and $\delta \chi^d$ stand for the continuous and discrete spectrum of linear modes, respectively.
It is natural to further decompose them into the real and imaginary parts, namely for the real part we have
\begin{align}
\label{solution_attra}
{\rm Re} \left[\delta \chi^c (x,t)\right]  &= \int_{-\infty}^{\infty} \mathrm{d} k \mathcal{A}^+_k (t) \varphi^+_k(x), \\
{\rm Re} \left[\delta \chi^d (x,t) \right] &= c_1(t)\gamma_1(x)+\tilde{c}_2(t)\tilde{\gamma}_2(x),
\end{align}
and similarly for the imaginary part
\begin{align}
\label{solution_attra1}
{\rm Im} \left[\delta \chi^c (x,t) \right] &= \int_{-\infty}^{\infty} \mathrm{d} k \mathcal{A}^-_k (t) \varphi^-_k(x),\\
{\rm Im} \left[\delta \chi^d (x,t)\right]  &= c_2(t)\gamma_2(x)+\tilde{c}_1(t)\tilde{\gamma}_1(x).
\end{align}
The explicit expressions for the complete spectrum of modes are given in Appendix~\ref{BdGA}.
Time dependence of the expansion coefficients of the continuous part takes a simple form
\be 
\mathcal{A}^{\pm}_k(t) = \cos(\omega_k t)\mathcal{A}^{\pm}_k(0) \pm \sin(\omega_k t) \mathcal{A}^{\mp}_k (0),
\ee
with the dispersion 
\be 
\omega_k = k^2 + \nu^2.
\label{dispersion_attr}
\ee
The dependence of the discrete part is at most linear in time
\be 
\tilde{c}_j(t)=\tilde{c}_j(0),\quad c_j(t) =c_j(t) + 2\nu \tilde{c}_j(0) t,\quad j = 1,2.
\ee
The initial values $\mathcal{A}^{\pm}_k(0)$, $c_{1,2}(0)$ and $\tilde{c}_{1,2}(0)$ are determined from the initial profile. 

Note that the presence of a soliton (nonzero $\nu$) opens a gap in the continuous spectrum of Bogoliubov quasiparticles~\eqref{dispersion_attr}. Moreover, the dispersion~\eqref{dispersion_attr} coincides with magnon excitations (linear spin waves) in the presence of a magnetic field ($h=\nu^{2}$). This is can be explained by the gauge similarity of the focusing NLS to the isotopic Landau--Lifshitz ferromagnet \cite{Faddeev}, and the fact that the field intensity $|\psi(x,t)|^2$ corresponds to the gradient of spin field.

We wish to stress that the discrete modes are essential to satisfy the completeness of the solutions of BdG equation~\eqref{linearized2}. They can be obtained, in particular, by the observation that if $\psi_s(x,t;n)$ represents a solution
of the 1D NLSE~\eqref{NLSE_attractive} depending on some parameter $n$, then the parametric derivative
\be
\delta_n\psi(x,t) \equiv \partial_n \psi_s(x,t;n),
\ee
 satisfies BdG equation~\eqref{bdg} (see also Refs. \cite{walczak2011exact, Dziarmaga_2004}). In our case, $n$ can be any parameter in the one-soliton solution \eqref{bright_soliton}, e.g. $u$, $v$, $\phi_0$, etc. Therefore, the discrete modes for $\delta \chi^d(x,t)$ are proportional to
$e^{-i \nu^2 t} \partial_n \psi_s(x,t)$ which corresponds to the discrete zero-energy solutions (also known as zero-modes).
Notice however that the discrete modes generated this way are not all linearly independent.
In Appendix~\ref{BdGA} we carefully check the completeness and orthogonality relations. 

The initial profile for BdG equation~\eqref{bdg} is obtained by subtracting the
soliton~\eqref{psiS} from the quenched profile~\eqref{initialAtrractive},
\be
\begin{split}
	\delta \psi(x,0) & = \psi(x,0) - \psi_s(x,0) = \\
	& =  \frac{1}{\sqrt{|\varkappa|}} \sech\left( \frac{x}{\eta} \right) - \frac{\nu}{\sqrt{|\varkappa |}} \sech(\nu x) .
\end{split}
\label{initial_bdg_attrac}
\ee
The initial values of the expansion coefficients of both continuous and discrete modes are calculated numerically from this profile.
For the discrete modes we find that $c_1(0) = c_2(0) = \tilde{c}_1(0) = 0$, while $\tilde{c}_2(0) \neq 0$, which means that 
the imaginary part of the evolved profile grows linearly with time. This, in particular,
results in the divergence of the conserved charges of the original nonlinear problem as $t \to \infty$, signaling a breakdown
of the linearization approach and its inadequacy for describing nonequilibrium scenarios such as the quenches presented here.

\begin{figure*}[b]
	\subfloat{\includegraphics[width = .5\linewidth]{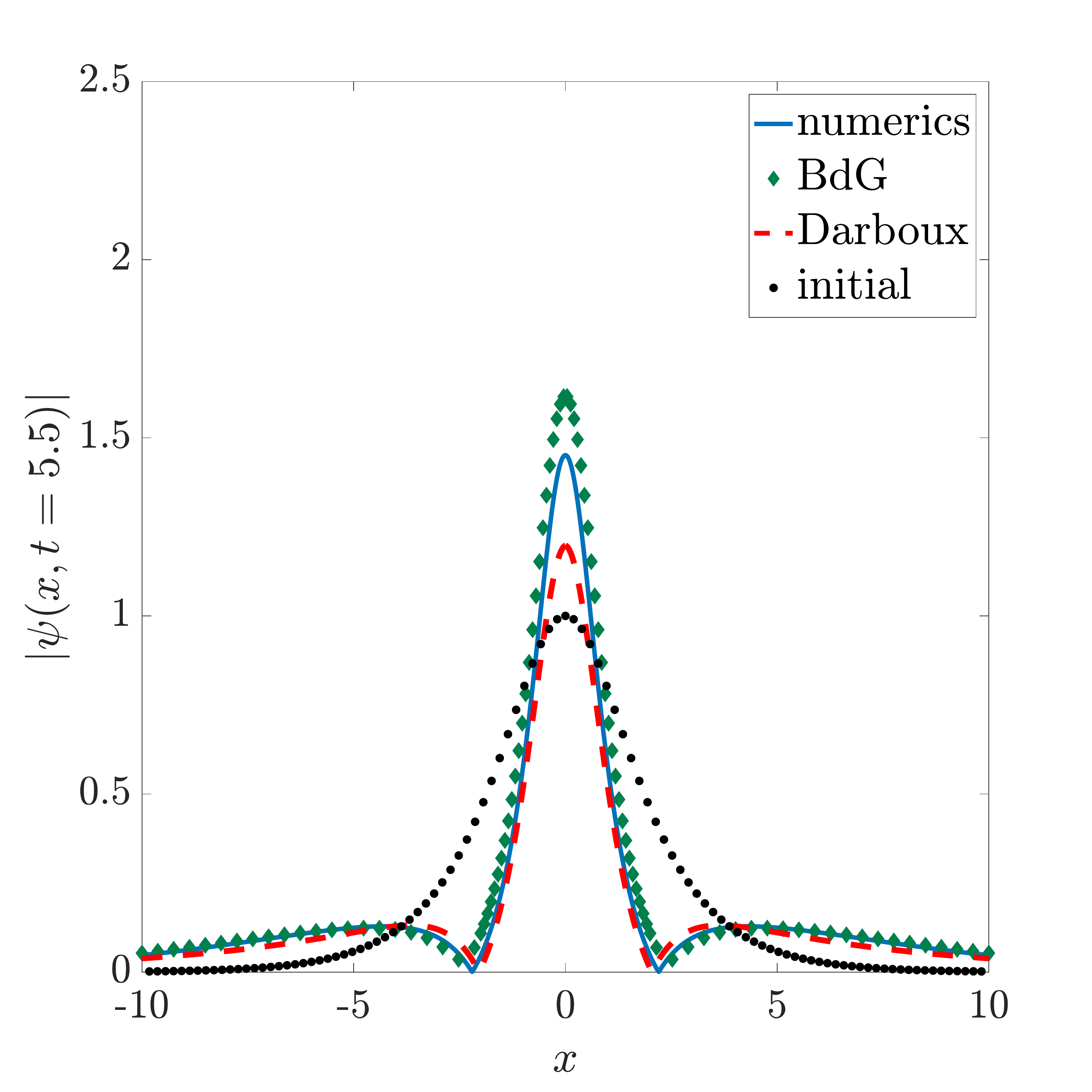}} 
	\subfloat{\includegraphics[width = .5\linewidth]{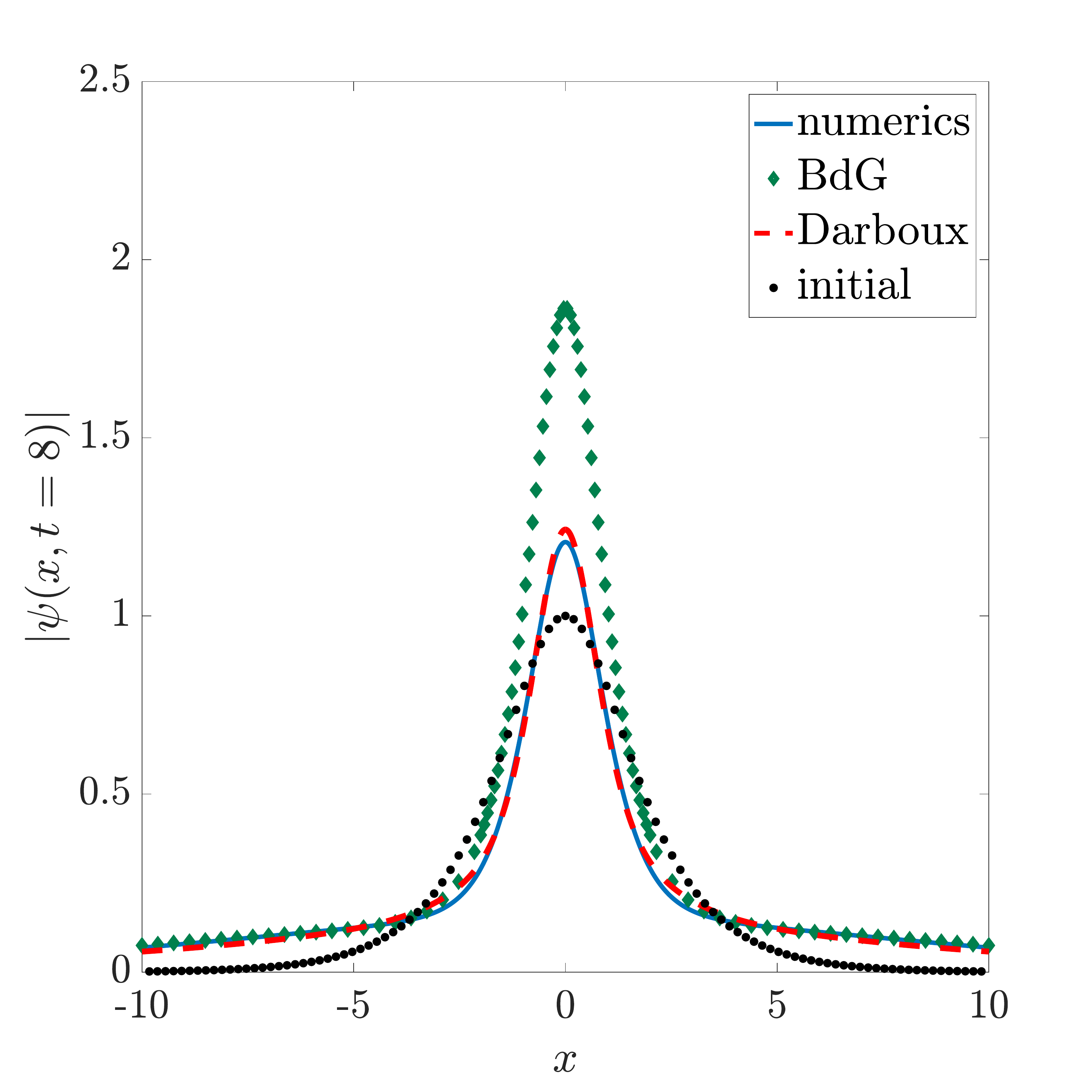}}\\
	\subfloat{\includegraphics[width = .5\linewidth]{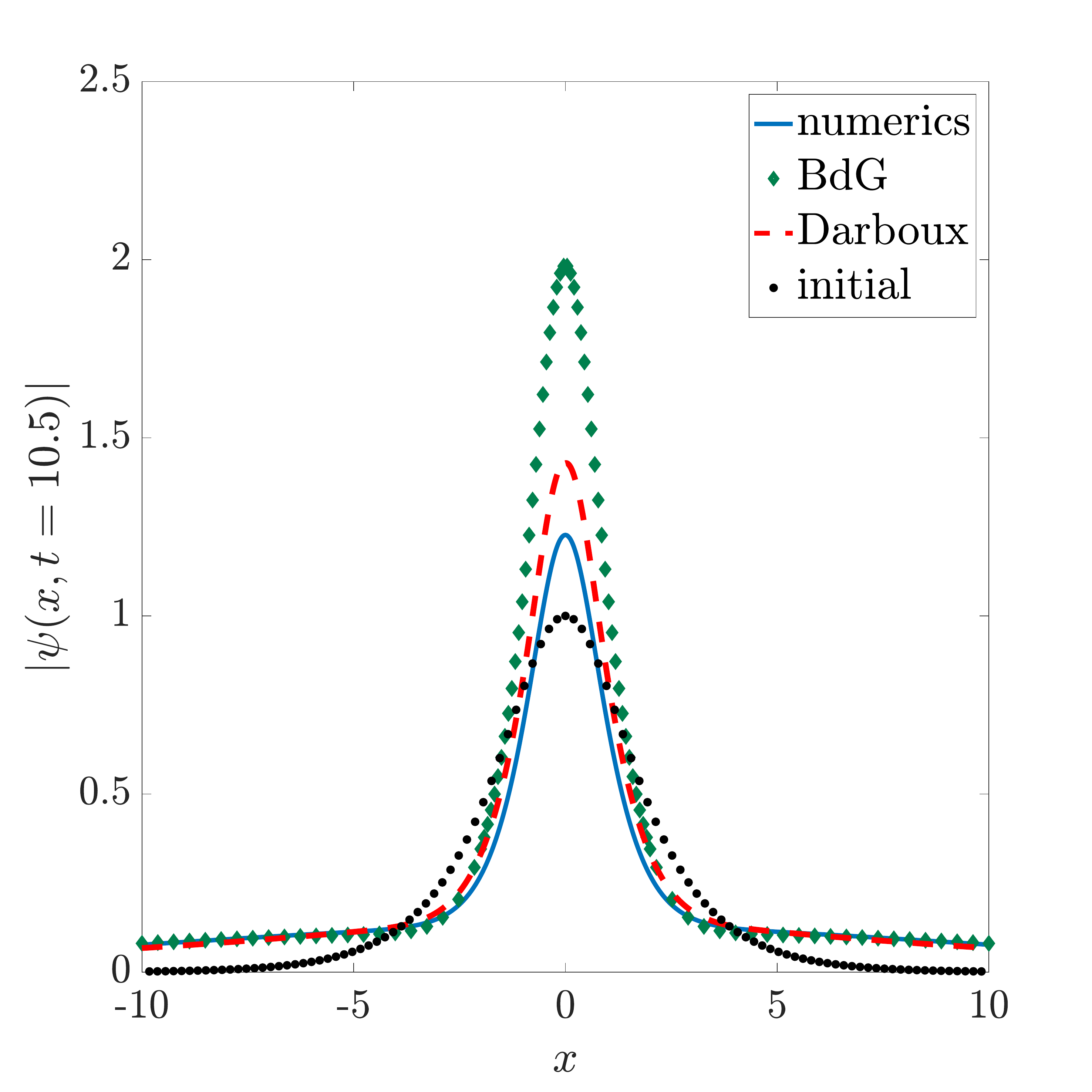}}
	\subfloat{\includegraphics[width = .5\linewidth]{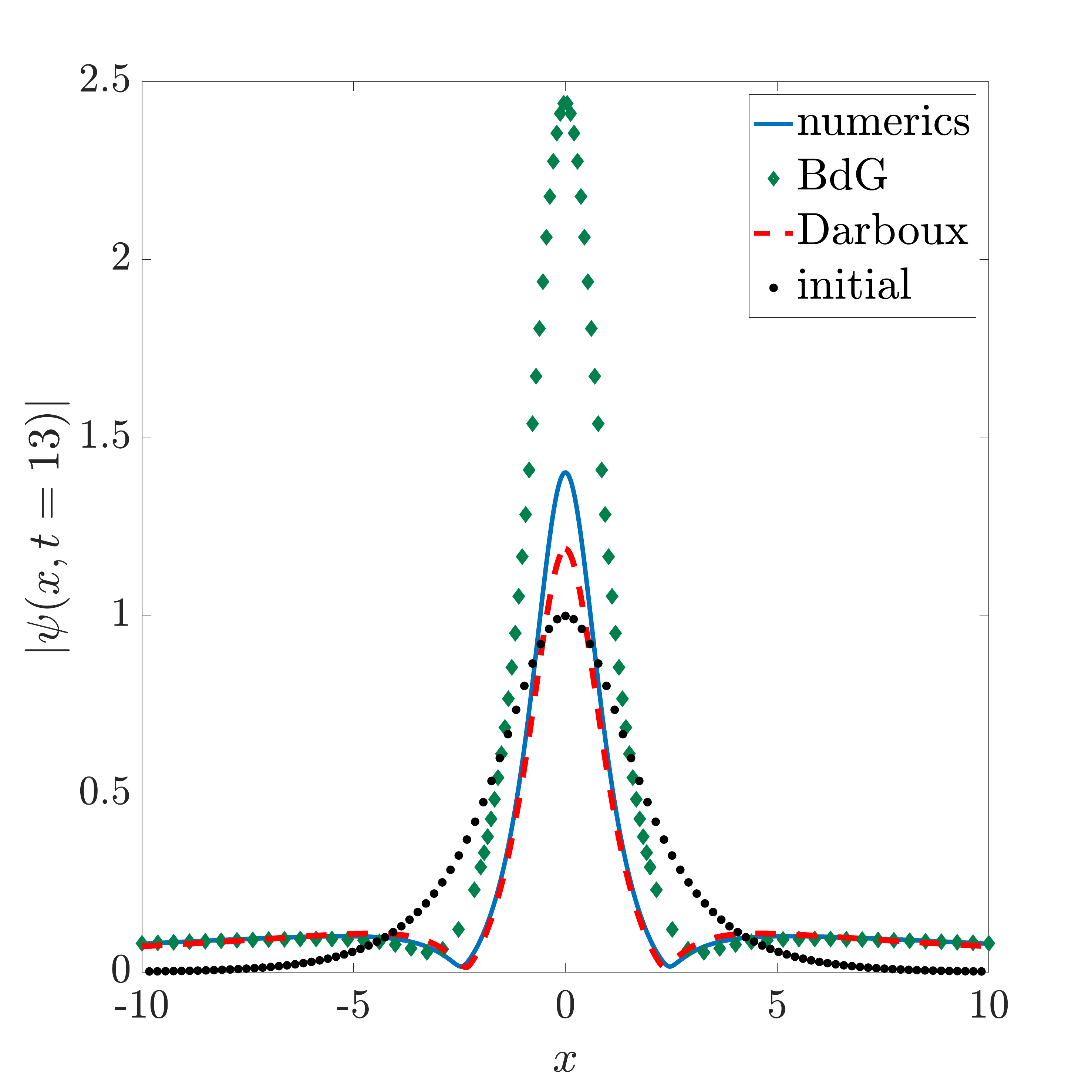}} 
	\caption{Solution to BdG equation~\eqref{linearized2} (diamonds) versus the solution constructed
	from the asymptotics profile dressed by a soliton~\eqref{bright_asymp_full} (dashed), shown for $\eta = \sqrt{2}$
	($\varkappa =- 1$) at times $t=5.5$, $8$, $10.5$, and $13$, respectively. The results are benchmarked against the numerical 
	integration (solid line).}
	\label{fig:Darboux}
\end{figure*}

In Fig.~\ref{fig:Darboux} we present a comparison with the exact numerics, demonstrating that in relatively short amount of time 
the discrepancy blows up. It is worth mentioning at this stage that a general analysis of nonlinear equations containing
soliton-like solutions suggests that the zero-mode contributions are absent in the asymptotics \cite{Buslaev1995}. Hence,
the results of the BdG theory can be trusted only on sufficiently short time-scales. Curiously enough, if we voluntarily discard the contributions of discrete modes, i.e. putting $\tilde{c}_2=0$ and retaining only the finite-frequency
continuum, the approximation improves quite noticeably, see Fig.~\eqref{fig:half_width_comp}.
In order to describe the large-time asymptotics, one has to be able to properly take into account the infinite set of the conserved charges and find their values on the initial profile. We achieve this in the next section using the dressing method.

\paragraph*{Darboux transformation.}


The conserved charges can be found using the asymptotic expansion \eqref{consvA} and the usual parametrization \eqref{a(l)BR}. Namely, they can be split into soliton and radiation contributions
\be 
Q_n = Q_n^{(s)} + Q_n^{(r)},
\label{QQ1}
\ee
where for $1/2<\eta <3/2$ one can easily deduce that  
\be 
Q_n^{(s)}  = i \frac{\lambda_1^n-\bar{\lambda}_1^n}{\varkappa n},\,\,\,\,\,\, \lambda_1 = i\nu,
\ee
whereas the radiation part $Q_n^{(r)}$ remains the same as in the solitonless case \eqref{QN1} (with the corresponding $|a(\lambda)|$ not depending on the soliton parameters). Hence, if we were only to include the constraints of $Q_n^{(r)}$, we would have arrived
at the same solution as previously, cf. Eq.~\eqref{bright_asymp_full}. To overcome this issue with only
minor modifications, we propose to employ the Darboux (dressing) transformation.

The very idea of the dressing transformation~\cite{Matveev_book} is to start from some reference solution and construct new solutions to the equation of motion by constricting a suitable nonlinear transformation. This will ensure that \eqref{QQ1} is satisfied exactly~\cite{gu2006darboux, its1988exact}. 
In distinction to the linearization procedure outlined previously, the present construct heavily relies on the underlying integrability, in particular on the existence of the auxiliary linear problem \eqref{ALP}, which in the short hand notations can be presented as
\be \label{u2}
\partial_x F = U F,
\ee
with connection
\be
\label{u}
U = \lambda \sigma_3/2i + U_{\psi},
\ee
and with $\sigma_3$ and $U_{\psi}$ defined in Eq.~\eqref{upsi}.
Applying a gauge transformation with the matrix $D = D(x,\lambda)$,
\be 
F \mapsto F^{\rm dr} = D F,
\ee
we find that $F^{\rm dr}$ satisfies the linear problem \eqref{u2} with the dressed connection
\be 
U \mapsto U^{\rm dr} = D U D^{-1} + \partial_xD D^{-1}.
\ee
We should in addition demand that $D$ is such that $U^{\rm dr}$ has the same structure as Eq.~\eqref{u}, but with the new field $\psi^{\rm dr}$ corresponding to another solution of the NLSE. The gauge $D$ can be sought as a polynomial in the spectral parameter $\lambda$, but for our purposes it is enough to consider a linear function
\be 
D = \lambda \mathbbm{1}- S(x).
\ee
The equation for $S$ then reads
\be  \label{S}
\partial_xS = [U_{\psi}+\tfrac{1}{2i}\sigma_3 S ,S],
\ee
while the dressed potential (connection) takes the form
\be\label{upsi2}
U^{\rm dr}_{\psi} =U_{\psi} +\tfrac{1}{2i}[\sigma_3,S].
\ee
To solve \eqref{S}, let $\mathcal{F} (\lambda_1)=(\mathcal{F}_1,\mathcal{F}_2)^T$ denote a general solution of the linear problem~\eqref{ALP} with $\lambda=\lambda_1$ and the potential of $U$ being $\psi(x,t)$. A formal expression is given by Eq.~\eqref{F1F2}. The solution for the complex-conjugate value $\lambda = \bar{\lambda}_1$ is given by $\mathcal{F}(\bar{\lambda}_1)=(-\bar{\mathcal{F}}_2, \bar{\mathcal{F}}_1)^T$. These two solutions can be combined in the matrix $ G = (\mathcal{F}(\lambda_1),\mathcal{F}(\bar{\lambda}_1))$, satisfying
\be 
\partial_x G = \frac{\sigma_3}{2i}G \Lambda +   U_{\psi} G, 
\ee 
with $\Lambda =  \frac{1}{2 i} {\rm diag}(\lambda_1,\bar{\lambda}_1)$. This way, the solution to Eq.~\eqref{S} is given by
\be 
S = G \Lambda G^{-1}.
\ee
The dressed field (cf. Eqs.~\eqref{upsi2}, and \eqref{upsi}) takes the form
\be \label{dr}
\psi^{\rm dr} = \psi  + \frac{\bar{\lambda}_1-\lambda_1}{\sqrt{|\varkappa|}} \frac{\mathcal{F}_1 \bar{\mathcal{F}}_2}{|\mathcal{F}_1|^2+|\mathcal{F}_2|^2}.
\ee
The scattering data $a^{\rm dr}(\lambda)$ for the dressed potential $\psi^{\rm dr}$ can be computed from the asymptotics of
$F^{\rm dr}$, which is determined by the asymptotics of $F$ (the undressed scattering data), and by the asymptotics 
of $S(x)$, which for ${\rm Im} \lambda_1 >0$ reads
\be 
\begin{split}
S(x\to+\infty) & = \frac{1}{2 i} \bp \lambda_1 & 0 \\ 0 &  \bar{\lambda}_1 \ep  ,\\ 
S(x\to-\infty) & =  \frac{1}{2 i} \bp \bar{\lambda}_1 & 0 \\ 0  & \lambda_1 \ep  .
\end{split}
\ee
Finally, the dressed scattering data is given by
\be \label{tildea}
a(\lambda) \mapsto a^{\rm dr}(\lambda) = \frac{\lambda-\lambda_1}{\lambda-\bar{\lambda}_1} a(\lambda).
\ee
The obtained expression is consistent with the general form \eqref{a(l)BR}, and moreover manifestly respects the form
of Eq.~\eqref{QQ1} for the conserved charges. In other words, we have constructed a new solution $\psi^{\rm dr}(x,t)$ to the equation 
of motion, which contains the original field $\psi(x,t)$ (described by $a(\lambda)$ in the spectral space) dressed by a soliton
(with the spectral parameter $\lambda_1$).

The most renowned application of the Darboux transformation is the construction of a one-soliton profile by dressing the
trivial vacuum solution $\psi(x) = 0$. In this case $a(\lambda)=1$, and one gets a one-soliton solution $\psi^{\rm dr}$ with
the corresponding $a^{\rm dr}(\lambda)$ given by Eqs.~\eqref{bright_soliton} and \eqref{asol}, respectively.
For the purposes of our application, we should instead apply the dressing to the asymptotic solution \eqref{bright_asymp_full}.
The condition \eqref{QQ1} is guaranteed to be satisfied exactly.
To get a symmetric profile we use a specific solution for $\mathcal{F}_{1,2}$ (see Eq.~\eqref{F12} and
appendix~\ref{app:attrac_app}). 

The comparison with numerical integration is shown in Fig. \eqref{fig:Darboux}. We see that the dressing method stays very 
close to the exact result for all times, similarly as in the solitonless case. The origin of a small discrepancy is due to the fact
that we have not used the exact solitonless solution but rather the asymptotic one given by Eq.~\eqref{F12}.

\section{Repulsive interaction}
\label{sec:dark}

Now we consider the repulsive interaction, also allowing for soliton modes if the system is initialized at finite density.
We consider the finite density repulsive (defocusing) NLSE,
\be
i \partial_t \psi(x,t) = - \partial_x^2 \psi(x,t) + 2 \varkappa (|\psi(x,t)|^2-\varrho)\psi(x,t) ,
\label{NLSE_replusive}
\ee
with nonlinearity coefficient $\varkappa>0$, and $\varrho$ denoting the asymptotic background density,
$|\psi(x)|^2 \to \rho$ as $x\to \pm \infty$.  With no loss of generality we subsequently put $\varrho = 1$.

The corresponding auxiliary linear problem is now of the form \cite{Faddeev}
\be 
\label{ALP2} 
\frac{{\rm d}}{{\rm d}x} \left(\begin{array}{c}
	F_1 \\
	F_2
\end{array}\right) =\left[\frac{\lambda}{2i}\sigma_3+ U^r_{\psi}\right]\left(\begin{array}{c}
F_1 \\
F_2
\end{array}\right).
\ee
with 
\be 
U^r_{\psi} = \sqrt{\varkappa} \left(\begin{array}{cc}
	0 & \bar{\psi} \\
	\psi & 0
\end{array}\right). 
\ee

As the initial profile we consider the ``rescaled dark soliton'' with quench parameter $\eta \in \mathbb{R}^+$ 
\be\label{psi0} 
\psi(x,0) = -\tanh (\sqrt{\varkappa}x/\eta ).
\ee
The corresponding scattering data can be computed directly from the Eq.~\eqref{ALP2},
reading~\cite{gamayun2015fate,gamayun2016soliton}
\begin{align}
a(\lambda) &= \frac{ik}{\lambda}\frac{\Gamma(-ik\eta / \zeta)^2}{\Gamma(-ik\eta / \zeta +\eta)\Gamma(-ik\eta / \zeta -\eta)}, \\
b(\lambda) &= \frac{\sin(\pi \eta)}{\sinh(\pi k \eta / \zeta)},
\end{align}
where $k=\sqrt{\lambda^2-\zeta^2}$ and $\zeta = 2 \sqrt{\varkappa}$. Parameter $\zeta$ is the speed of sound, corresponding
to the upper limit for the velocity of dark/gray solitons~\cite{Pitaevskii_2016,Faddeev}. The existence of such an upper bound is contrastingly different from the case of bright solitons in the 1D attractive NLSE~\eqref{NLSE_attractive} where the
the range of soliton velocities is not restricted. Another important difference with respect to the attractive case is that
for repulsive interactions the soliton part is present for all $\eta \in \mathbb{R}^+$. Recall that solitons correspond to zeros of $a(\lambda)$ in the upper-half spectral plane, so there are $2\lfloor \eta \rfloor +1$ solitons coexisting with the radiative part (which exactly vanishes for the integer $\eta$). In particular, for $0<\eta<1$ we can present
\be 
a(\lambda) = i\frac{k-i\zeta}{\lambda} \frac{\Gamma(-ik\eta / \zeta)\Gamma(-ik\eta / \zeta +1)}{\Gamma(-ik\eta / \zeta +\eta)\Gamma(-ik\eta / \zeta -\eta+1)},
\ee
or, alternatively, using the uniformization spectral variable, $z$ defined by
\be
\lambda = \frac{1}{2}\left(z+\frac{\zeta^2}{z}\right),\qquad k = \frac{1}{2}\left(z-\frac{\zeta^2}{z}\right),
\label{lambdaK}
\ee
in an equivalent canonical form
\be 
a(z) = i\frac{z - z_{1}}{z - \bar{z}_{1}} \exp\left[\int\frac{\mathrm{d} s}{2\pi i} \frac{\log (1+|b(s)|^2)}{s-z}\right].
\label{dark_a}
\ee
The soliton part now describes a dark soliton with spectral parameter $z_{1} = i \zeta$. 
The corresponding one-soliton profile is of the form 
\be 
\psi_s(x) = -\tanh (\zeta x/2).\label{solD}
\ee 
Similarly to the attractive case~(cf. Eq.~\eqref{a(l)BR}), the remaining analytic part of $a(z)$ can be seen as the radiation modes, characterized by nonzero reflection coefficient $r(z) = b(z)/a(z)$.

A physical consequence of repulsive interactions is that the radiation effectively decouples from the solitons.
This means that to obtain the asymptotic profile it will be sufficient to consider their contributions independently.
We proceed by computing the asymptotics of the generic solitonless profile, following
the same lines as in the attractive case (cf. Secs.~\ref{subsec:solitonless_bright}).
Subsequently we shall analyze the Bogoliubov theory describing linear fluctuations on the soliton background.
We show that the decoupling between the radiation and soliton components is quite profound and in fact happens exponentially fast.
The main benefit of this observation is that in order to describe asymptotic solutions in the presence of both the solitons
and radiation there is no longer any need of resorting to the Darboux transformation.

\subsection{Solitonless asymptotic}
\label{subsec:solitonless_dark}

Similarly to the attractive case, we consider linear perturbations of the trivial vacuum, which we choose to be $\psi^{(0)}(x)=1$.
We put $\psi(x,t) = 1+\delta \psi(x,t)$ to obtain from Eq.~\eqref{NLSE_replusive} a linear equation	
\be 
i\partial_t \delta\psi = - \partial_x^2 \delta\psi + 2\varkappa\,\delta\psi + 2\varkappa\,\delta \bar{\psi},
\ee 
which has a general solution of the form
\be \label{deltaPsi}
\delta \psi = \int \frac{\mathrm{d} k}{\sqrt{2\pi}} z(k)\left(C_+(k) e^{ig_+(k)}
+ C_-(k) e^{ig_-(k)} \right),
\ee
where
\be 
z(k) = k + \sqrt{k^2+\zeta^2},\,\,\,\,\,\,\, \omega(k) =  k \sqrt{k^2+\zeta^2 },
\label{disp_dark_no_soli}
\ee
\begin{figure}
	\centering
	\includegraphics[width=\linewidth]{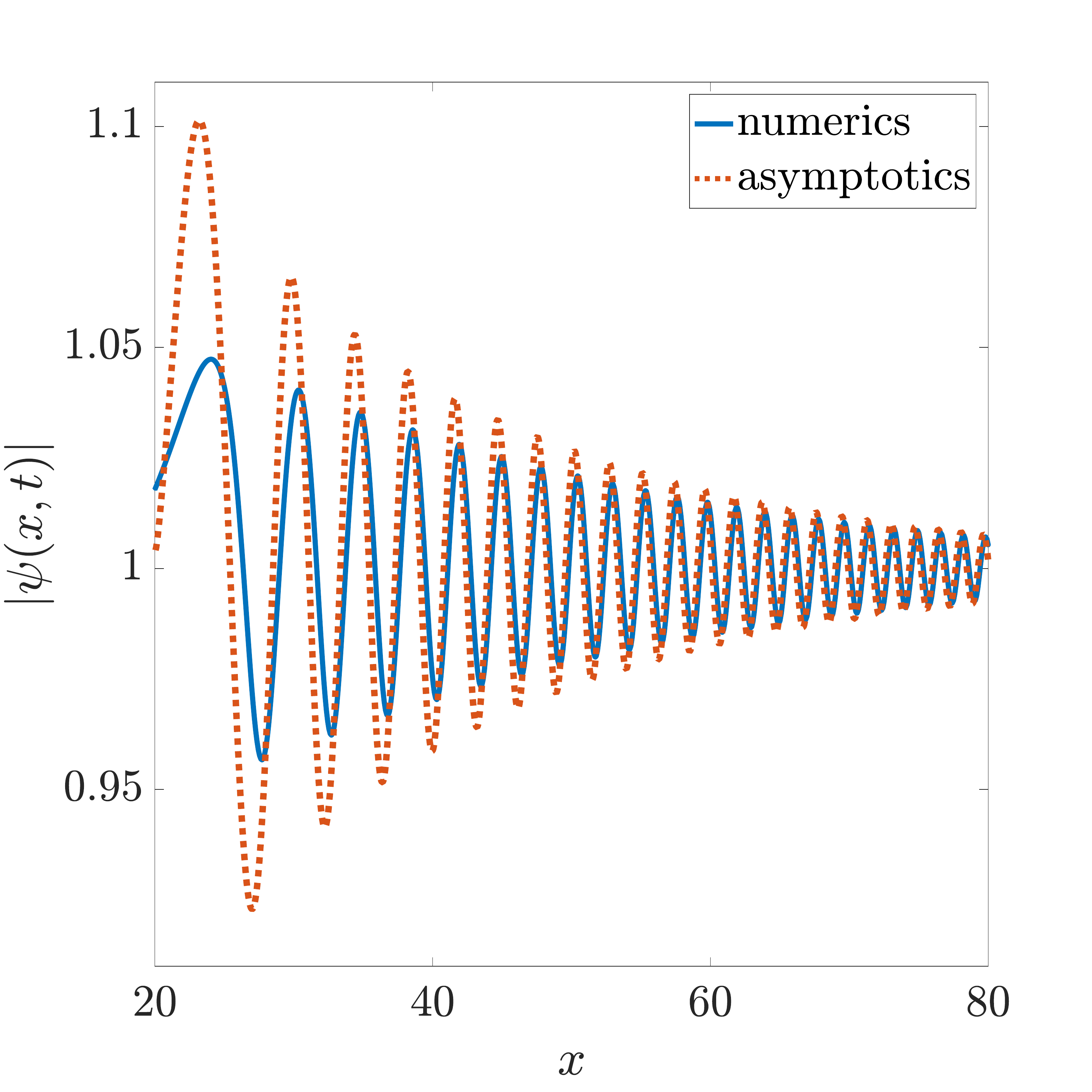}
	\caption{The asymptotic solution of Eq.~\eqref{ttyy} at $t=10$ (dotted line), compared with the numerical solution of the exact equation of motion~\eqref{NLSE_replusive} (solid line), for parameter $\varkappa = 1$ and $\eta = \frac{1}{2}$.}
	\label{fig:dark_asymp}
\end{figure}
\be 
g_{\pm} (k) = \pm  k x -  \omega(k) t,
\ee
with $C_\pm(k)$ being arbitrary functions with the following reflection property $C_\pm(-k) =-\bar{C}_\pm(k)$.
For a smooth initial profile, the $t\to +\infty$ asymptotic behavior of $\delta \psi(x,t)$, $\xi\equiv x/t = \mathcal{O}(1)$
can be calculated by the steepest descent method.
This requires to find the critical points $k_c$ of the phase factor $g_{\pm}(k)$, 
\be 
\frac{\mathrm{d} g_{\pm} (k)}{\mathrm{d} k}\Bigr\rvert_{k = k_c} = 0.
\ee
One can immediately notice that there are no solutions for $|x|< \zeta t$,
meaning that $\delta \psi$ is exponentially small in that region.
This provides an explanation for the supersonic nature of the radiation modes, as discussed later in Sec.~\ref{subsec:dark_linear}.
For $x> \zeta t$, the critical points $k=\pm k_c$ are contained only in the ``$C_+$ part'' of Eq.~\eqref{deltaPsi}.
To describe them explicitly we introduce $\lambda_c = \sqrt{k_c^2+\zeta^2}$ and $z\equiv z(k_c)=\lambda_c + k_c$, $z>\zeta$.
Then the $\xi$-dependence of the critical points can be found from the relation
\be 
\lambda_c = \sqrt{k_c^2+ \zeta^2} = \frac{\xi+\sqrt{\xi^2+8\zeta^2}}{4},\qquad \xi = x/t.
\label{lambda_c}
\ee
Notice also that $k_c$, $\omega_c$ and $\lambda_c$ have the same form as for a suitable parametrization for the scattering data \eqref{lambdaK}
\be 
k_c = \frac{1}{2}\left( z-\frac{\zeta^2}{z} \right),\quad \omega(k_c) = \frac{1}{4} \left( z^2 - \frac{\zeta^4}{z^2} \right) ,
\label{k_c}
\ee
however, in this case $z$ is not a spectral parameter but a function of $x$ and $t$. (cf. with $\mu$ and $x/2t$ for the attractive case \eqref{subsec:solitonless_bright}). 

The critical points correspond to
\be
\xi = \frac{z^4 + \zeta^4}{z(z^2+\zeta^2)} , 
\ee
which means that
\be 
\frac{{\rm d}\xi}{{\rm d}z} = \frac{(z^4 + 4 \zeta^2 z^2 + \zeta^4) (z^2 - \zeta^2)}{z^2( z^2+ \zeta^2)^2} .
\ee
This way, the asymptotics of $\delta \psi$ for $x>\zeta t$ reads
\be \label{ttyy}
\begin{split}
\delta \psi & = \frac{z C_+(k_c)e^{-i\pi/4}}{\sqrt{t \omega^{\prime \prime}(k_c)}} + \frac{\zeta^2 C_+(-k_c)e^{i\pi/4}}{z\sqrt{t \omega^{\prime \prime} (k_c)}} \\
 &  = \frac{\mathfrak{f}(z)}{\sqrt{z t(z+\zeta^2 /z)}}\sqrt{2\frac{{\rm d}z}{{\rm d}\xi}} \left(
z e^{i\theta} - \frac{\zeta^2 e^{-i\theta}}{z} \right) ,
\end{split}
\ee
where
\begin{align}
\omega^{\prime \prime}(k) &= \frac{2z}{z+ \zeta^2 /z} \frac{{\rm d}\xi}{{\rm d}z}, \\ 
\theta &= k_c x -\omega(k_c) t+{\arg } \, C_+(k_c) - \frac{\pi}{4},\\
\mathfrak{f}(z) &= |C_+(k_c)| \frac{(z+\zeta^2 /z)}{2}.
\end{align}

Similar expressions can be found for $x<-\zeta t$. Utilizing the same approach as in Eq.~\eqref{asymptBR}, we can write down
the following ansatz for the asymptotic expansion
\begin{align}
\label{dark_asymp}
	\delta \psi(x,t)& = \frac{\tilde{f}(z,t)}{\sqrt{z t(z+\zeta^2 /z)}}\sqrt{2\frac{dz}{d\xi}} \\
	&\times  \left( z e^{i\varphi(z,t)+i\delta\varphi(z,t)} - \frac{\zeta^2}{z} e^{-i\varphi(z,t)-i\delta\varphi(z,t)} \right),
	\nonumber
\end{align}
where
\begin{align}
\tilde{f} (z,t) &= f(z) + \sum_{n=1}^{\infty} \sum_{k=0}^{n} \frac{(\log t)^k}{t^n} f_{nk}(z),\\
\varphi (z,t) &= [ k_c(z) \xi(z) - \omega(z) ] t + \Xi(z) \log t,\\
\delta\varphi(z,t) &= \delta\varphi_0(z)+\sum_{n=1}^{\infty} \sum_{k=0}^{n} \frac{(\log t)^k}{t^n} \delta\varphi_{nk}(z) .
\end{align}
In analogy to the attractive case, we fix $|f(z)|$ by evaluating the local conserved charges on the asymptotic profile \eqref{deltaPsi} and matching it to the initial values obtained from the spectral presentation. The calculations are
slightly more involved and thus are relegated to Appendix~\eqref{app:rep_app}. The end result is however analogous to \eqref{fa},
reading
\be \label{fr}
\left| f(z) \right|^2 = \frac{1}{8 \pi \varkappa} \log(1+|b(z)|^2).
\ee
The logarithmic phase correction $\Xi(z)$ is obtained after substituting the asymptotic ansatz~\eqref{dark_asymp} into the
1D NLSE~\eqref{NLSE_replusive},
\be 
	\Xi(z) = \frac{z+\zeta^2 /z}{2\pi z} \frac{\mathrm{d} z}{\mathrm{d} \xi} \log(1+|b(z)|^2) .
\ee
This method allows us to fix the ``action part'' of the asymptotic solution and write
\begin{align}
\label{dark_asymp1}
\delta \psi(x,t) & = \frac{|f(z)|}{\sqrt{z t(z+\zeta^2 /z)}}\sqrt{2\frac{dz}{d\xi}} \\
&\times  \left( z e^{i\varphi(z,t)+i\delta\varphi_0(z)} - \frac{\zeta^2}{z} e^{-i\varphi(z)-i\delta\varphi_0(z)}\right).\nonumber
\end{align}
In the same manner as in the attactive case (cf.~\eqref{o1phase}) the phase of $\delta \varphi_0(z)$, encoding the angle variables \cite{A.R.Its1988}, can be fixed from the corresponding Riemann--Hilbert problem \cite{DeiftP.A.1993, A.R.Its1988,Vartanian_2000}.  

The comparison of asymptotic results against the numerics is displayed in Fig. \eqref{fig:dark_asymp}. There
we consider the $\eta=1/2$ quench which involves a dark soliton, but primarily focus on the region with no solitons.
Next we shall discuss the separation between radiation modes and solitons in more detail.

\subsection{Solitons versus radiation at finite density}
\label{subsec:dark_linear}

To gain some insight into how the presence of a soliton influences the analysis of the previous section, we follow the logic of the attractive case and consider a linearized theory with respect to the soliton background. Putting $\psi = \psi_s + \delta \psi$,
we obtain a linearized equation
\be \label{solL}
i\partial_t \delta \psi  = - \partial_x^2 \delta \psi + 2 \varkappa (2|\psi_s|^2-1 ) \delta \psi + 2 \varkappa \psi_s^2 \delta\bar{\psi},
\ee
describing the BdG equation of the following matrix form
\be 
	i \partial_t \bp \delta \psi \\ \delta \bar{\psi} \ep = \bp A & B \\  -\bar{B} & -A \ep \bp \delta \psi \\ \delta \bar{\psi} \ep.
\ee
Here $A = - \partial_x^2 + 4 \varkappa  |\psi_s |^2 - 2\varkappa $, $B = 2\varkappa \psi^2_s$ and the soliton solution $\psi_s$ is given by \eqref{solD}. Unlike in the vacuum case \eqref{deltaPsi},
the general solution of Eq.~\eqref{solL} now comprises both the continuous and discrete modes.
\begin{figure}
	\centering
	\includegraphics[width=\linewidth]{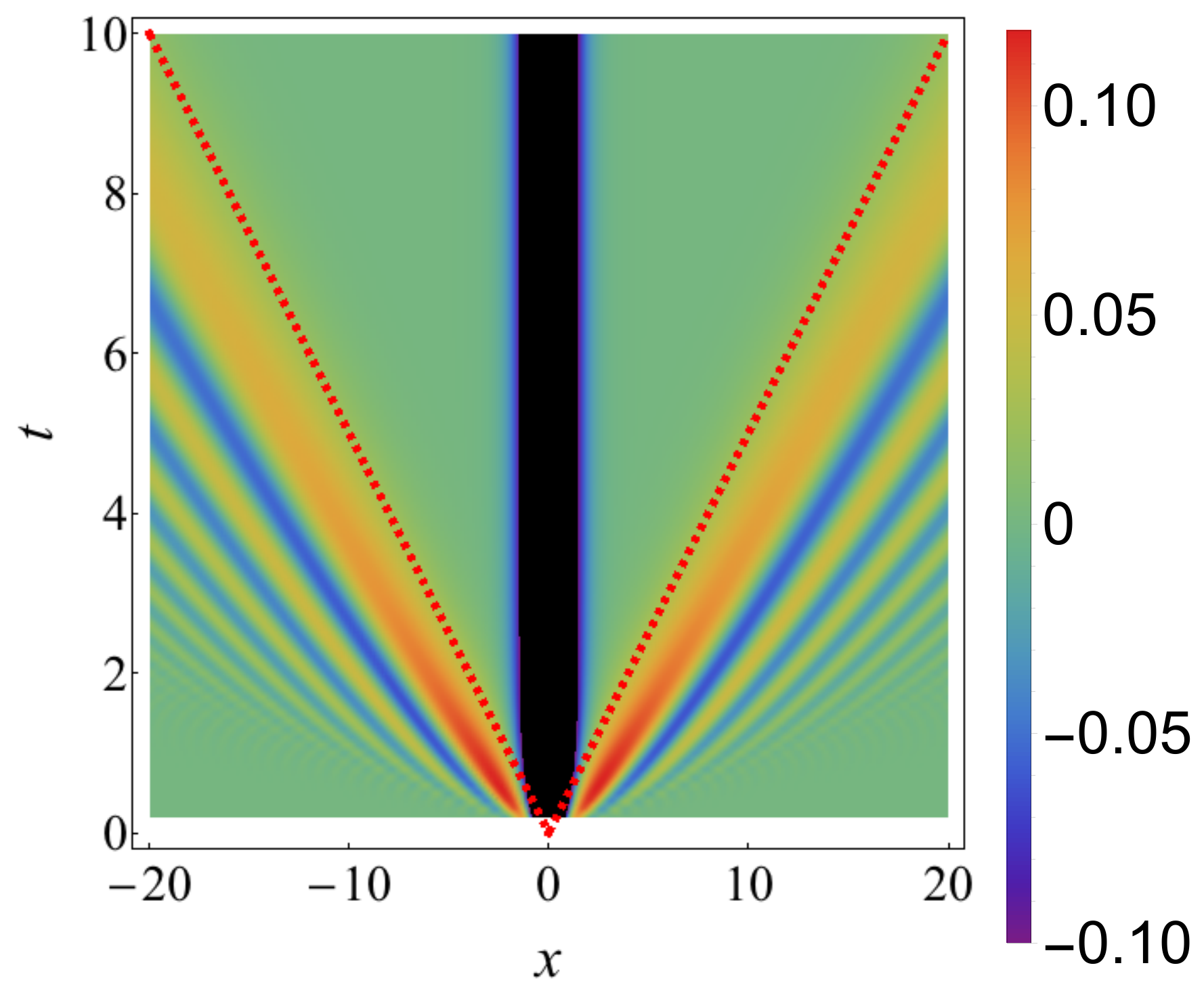}
	\caption{Time evolution of $|\psi(x,t)|-1$ of the quenched profile~\eqref{psi0} with coupling $\varkappa=1$ and quench parameter $\eta = 1/2$. The radiation modes propagate faster than any dark/gray solitons, with lower velocity bound $\zeta$. All the radiation modes are confined outside of the sound cone, marked with dashed lines.}
	\label{fig:anti_lightcone}
\end{figure}
Formally, it can be split as
\be \label{generPsi}
\delta \psi = \delta \psi^c + \delta \psi^d
\ee
where the continuous part $\delta \psi^c$ is further decomposed into normal modes, which for convenience we present separately
for real and imaginary parts,
\begin{align}
\hspace{-2mm}{\rm Re}\left[\delta \psi^c(x,t)\right] &= \int\limits_0^\infty \!{\rm d}k \left[\mathcal{A}_k^+(t)\varphi_k^+(x) +\bar{\mathcal{A}}_k^+(t) \phi^+_k(x) \right], \\
\hspace{-2mm}{\rm Im}\left[\delta \psi^c(x,t)\right] &= \int\limits_0^\infty \!{\rm d}k \left[\mathcal{A}_k^-(t)\varphi_k^-(x) +\bar{\mathcal{A}}_k^-(t) \phi^-_k(x) \right].
\end{align}
The exact form of $\varphi^\pm$ and $\phi^\pm$, as well as their orthogonality relations, are presented in Appendix \eqref{app:repulsive_solu}. The time dependence of the expansion coefficients is given by
\be 
	\mathcal{A}_k^{\pm} (t) = \cos (\omega_k t) \mathcal{A}^{\pm}_k (0) \pm  \sin (\omega_k t) \mathcal{A}^{\mp}_k (0),
\ee
with the dispersion relation
\be 
	\omega_k = k \sqrt{k^2+\zeta^2} .
	\label{dispersion_dark}
\ee
One can observe that this dispersion is exactly the same as for excitations on the solitonless background.
For small momenta it describes linear sound excitations, indeed $\omega(k) \approx \zeta k$ as $k\to 0$.
Notice that the group velocity of propagation always exceeds the sound velocity $\zeta$
\be 
v_r = \frac{\mathrm{d} \omega_k }{\mathrm{d} k} = \sqrt{k^2 + \zeta^2} + \frac{k^2}{\sqrt{k^2+ \zeta^2}} \geq  \zeta ,
\ee
meaning that the radiation modes are ``supersonic'', while the solitons on the contrary have
``subsonic'' velocities $|v_{\rm sol}|<\zeta$. Therefore, irrespective of the initial profile, the overlap between 
radiation and solitons becomes exponentially small in a short amount of time.
This can be clearly observed in the numerical simulations of the exact nonlinear dynamics of 1D NLSE~\eqref{NLSE_replusive} 
shown in Fig.~\ref{fig:anti_lightcone}. We would like to emphasize that
all these effects can be attributed to the finite density boundary condition which
(irrespective of the sign of the coupling constant) changes the dispersion relation of the Bogoliubov quasi-particles from
the magnonic (i.e. quadratic) relation \eqref{dispersion_attr} to \eqref{dispersion_dark}, with linear dependence at long wave-lengths 
as for acoustic phonons.

In addition to the continuum part, we have the discrete modes $\delta \psi^d(x,t)$, which still can be regarded as zero-modes even though there is no gap in the spectrum. Their general form reads
\be 
\delta \psi^d = (c_1 - 2 \zeta c_2 t ) \frac{\zeta}{4} \sech^2 \left( \frac{\zeta x}{2} \right)  + i c_2 , 
\ee
with real constants $c_1$ and $ c_2$. These modes are essential to prove the completeness relations discussed
in Appendix~\ref{app:repulsive_solu} (cf. Eqs.~\eqref{ortho_dark1}, \eqref{ortho_dark2} and \eqref{complR}).

\begin{figure}
	\centering
	\includegraphics[width=\linewidth]{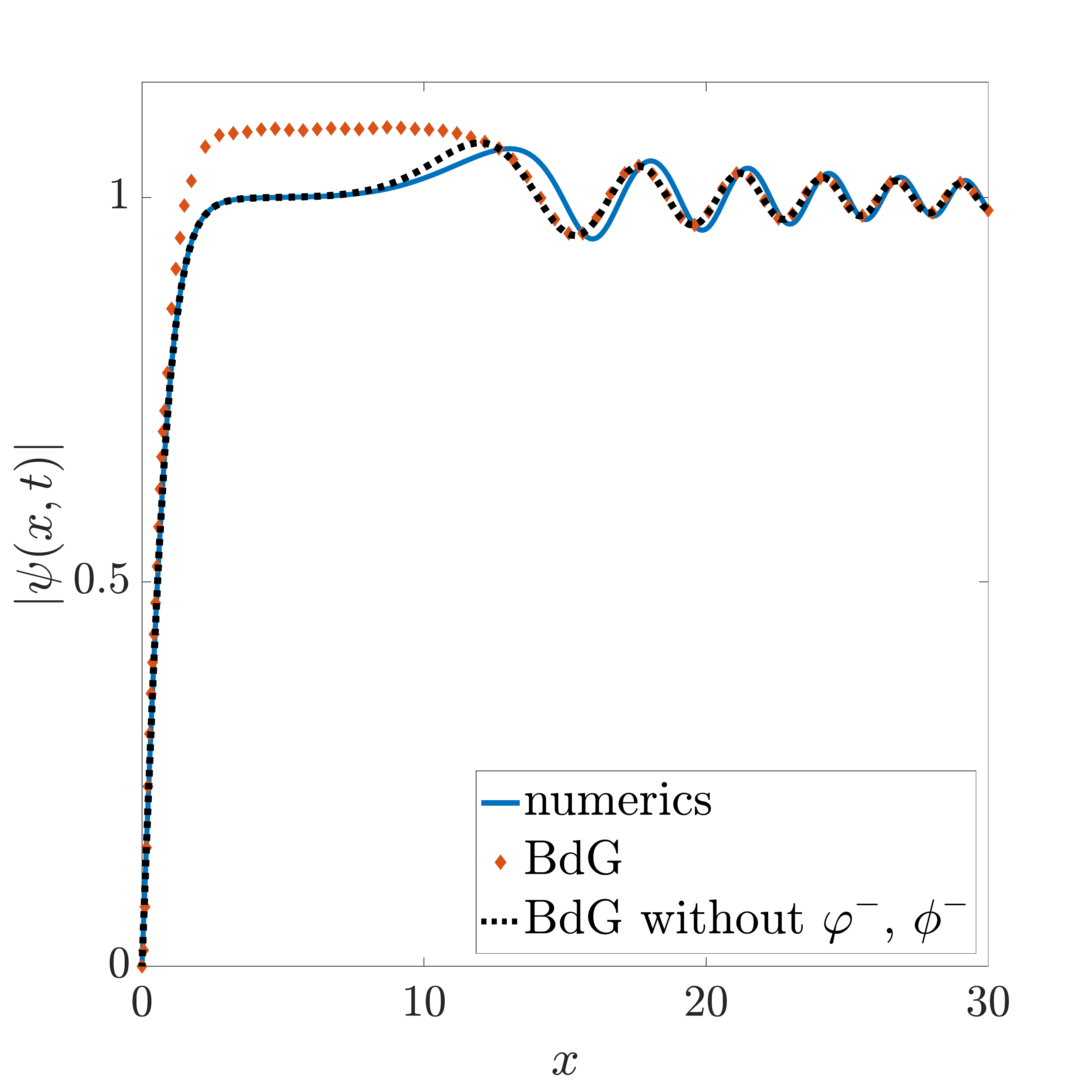}
	\caption{Solution of BdG equation~\eqref{solL} (diamonds) compared to the exact numerical integration (solid line)
	 for the 1D NLSE with finite asymptotic density~\eqref{NLSE_replusive}, with parameters $\eta = 1/2$ at $t = 5$. The imaginary part of the solution~($\varphi^{-}$ and $\phi^{-}$ modes) is responsible for the observed excess in the asymptotic density
	 (as shown by the dotted line).}
	\label{fig:dark_BdG}
\end{figure}
The initial values of $c_1$ and $c_2$, as well as, $\mathcal{A}^{\pm}_k (0)$ and $\mathcal{B}^{\pm}_k (0)$ are determined from the initial condition, which for the quench~\eqref{psi0} can be written as
\be 
\delta \psi(x,t=0) = \tanh\left(\frac{\zeta x}{2}\right) - \tanh\left(\frac{\zeta x}{2\eta}\right).
\label{init_dark}
\ee
This profile is antisymmetric which implies no contribution from the zero-modes, $c_1=c_2=0$, meaning that
the linear evolution is governed entirely by the quasi-particle continuum.

The linearized evolution is compared with the exact numerics in Fig. \eqref{fig:dark_BdG}, where the agreement at
sufficiently large distances is quite good.
It is particularly interesting to notice that the imaginary part (i.e. the contributions of $\varphi^-_k$ and $\phi^-_k$) results in a 
higher asymptotic density near the origin (see red curve in Fig.~\ref{fig:dark_BdG}). This happens because the full density 
($\rho(x)=|\psi_s+\delta \psi|^2 - 1$) is not a conserved quantity of the linearized BdG equation~\eqref{solL}.
However, if we again permit ourselves to discard this contribution, we get a much better approximation (see dashed black curve
in Fig.~\ref{fig:dark_BdG}).

For a generic profile with $c_2 \neq 0$, the BdG approach is a reasonable approximation only for short times. As commented
earlier, unlike in the attractive case now there is not need of performing a Darboux transformation to merge the solitons and 
radiation. Indeed, due the to the exponentially small overlap the conserved charges separate automatically
\be 
Q_n = Q_n^{(s)} + Q_n^{(r)},
\ee
and produce correct charges, modulo terms which become vanishing at large times.
This way, one can represent the asymptotics as a sum of the solitons (which can be read off from the spectral data) and
the asymptotic solution for the radiation~\eqref{dark_asymp1}.

\section{\label{sec:conc}Conclusions}

To conclude, we have studied the time evolution of the quenched soliton profiles in the classical
one-dimensional nonlinear Schr\"{o}dinger equation, both in the attractive and repulsive regimes.
We developed an analytic technique which combines the exact asymptotic formulae and the so-called Darboux dressing transformation, 
allowing us to accurately approximate the relaxation dynamics on large spatio-temporal scales.
In practice, our method yields satisfactory results even for moderately short times.

We first considered the attractive interaction and studied the time evolution of a solitonless initial profile.
We demonstrated that the intensity of the asymptotic field is completely determined by the initial values of the local conservation, 
bearing some conceptual similarity with the widely studied local equilibration in nonergodic quantum gases.
If one further chooses to discard the information about phases, e.g. by performing a uniform sampling over dynamical phases for an
ensemble of initial conditions, individual field configurations become locally indistinguishable and one can think of
the asymptotics formula as the classical (zero-density) analogue of the (generalized) Eigenstate Thermalization Hypothesis.

We subsequently studied the case when the initial profile contains solitons. The general conclusion is that
the linearized Bogoliubov--de~Gennes equations are inadequate, resulting in an unphysical diverging asymptotics
attributed to the presence of the zero-modes which involve terms growing linearly in time.
This pathology can be remedied by exploiting the underlying integrability and performing an exact dressing (Darboux) transformation,
permitting to superimpose a soliton mode on a radiative background and thus ensuring manifest preservation of the integrals
of motion. As demonstrated, this correctly explains the breathing of solitons which results from the interplay between the
solitons and radiation.

By repeating a similar analysis for the quenches in the repulsive case, we found a physical picture which radically differs
from the attractive regime. Owing to the fact that the soliton-radiation interactions get exponentially
suppressed with time, it is permissible to regard them as independent components. In particular, the quenched initial
profile emits radiation in the form of an undulating wave which propagates at supersonic velocity.

The methods presented in this manuscript are generic and can be applied to other exactly solvable classical field
theories or lattice models. One may, for instance, conceive similar inhomogeneous quenches in integrable classical models
which exhibit relativistic invariance or topological excitations (e.g. sine-Gordon equation),
which might unveil additional features beyond the phenomena covered in this paper.
Besides that, there are several other theoretical aspects which deserve to be addressed in future works.
An interesting question regarding relaxation in integrable classical soliton systems at finite energy density
and the emergence of local equilibrium ensembles still remains to be understood. To study transport phenomena,
it would be useful to translate and adapt the ideas of the recently developed “generalized hydrodynamics”
developed for quantum integrable systems~\cite{Castro_Alvaredo_2016, Bertini_2016} to the classical soliton theories,
as e.g. outlined in~\cite{Bastianello_2018}, or by identifying the thermodynamic limit of adiabatically modulated finite-gap solutions~\cite{Enej_Sasha_to_appear}. 

We expect that the theoretical protocol studied in our work can be realized using modern experimental techniques with ultracold atoms which, among other, enable to create soliton-like excitations in 1D Bose gases in both attractive and repulsive cases~\cite{lepoutre2016production}.
While the present study is restricted to study the the nonequilibrium dynamics of one-dimensional Bose gases 
using the classical (mean-field) description, in the actual experimental realizations with cold atom the presence of the confining 
potentials~\cite{PhysRevA.66.063602} or the many-body effects~\cite{PhysRevLett.89.200404, PhysRevLett.100.130401, PhysRevLett.119.220401, PhysRevA.96.043616, 2018arXiv181106612L} may also play an important role. It would be valuable to further 
investigate these matters in the future work.
\\

\begin{acknowledgements}
We thank Jean-S{\'e}bastien Caux for innumerable discussions and critical remarks on the manuscript,
Oleg Lychkovskiy for careful reading of the manuscript and Boris Malomed for drawing our attention to Ref. \cite{Malomed_1987}. Y. M. thanks Neil Robinson for useful discussions and suggesting the open-source solver ``\textit{chebfun}''~\cite{Driscoll2014}, and Giuseppe Mussardo for showing interest in our work. Y. M. and O. G. acknowledge the support from the European Research Council under ERC Advanced grant 743032 DYNAMINT. E. I. is supported by VENI grant number 680-47-454 by the Netherlands Organisation for Scientific Research (NWO).
\end{acknowledgements}

\onecolumngrid

\appendix

\section{\label{app:attrac_app}Classical Inverse Scattering Method for 1D Attractive NLSE}

In this appendix we describe a formal solution of the linear system \eqref{ALP}. 
This is needed to obtain the generating functions of the conserved charges and
to compute the Darboux transformation.
Following Ref. \cite{Ablowitz_1981}, the linear system \eqref{ALP}, reading
\be\label{ALPA}
\partial_x \bp F_1 \\ F_2 \ep = \left[\frac{\lambda}{2i}\left(\begin{array}{cc}
	1& 0\\
	0 & -1
\end{array}\right) + \sqrt{|\varkappa |} \left(\begin{array}{cc}
	0 & i\bar{\psi} \\
	i\psi & 0
\end{array}\right) \right] \bp F_1 \\ F_2 \ep,
\ee
can be reduced to the Ricatti equation. Namely, we first parametrize $F_1$ in terms of function $f(x,\lambda)$,
\be 
F_1 = \exp\left(\frac{\lambda}{2i} x +  f(x,\lambda)\right).
\ee
Then, using the first row in \eqref{ALPA}, we can express $F_2$:
\be 
F_2 = - \frac{\sqrt{|\varkappa|} \rho}{\bar{\psi} }  \exp\left(\frac{\lambda}{2i} x + f(x,\lambda) \right),
\ee
where we have introduced $\rho = i \partial_x f / |\varkappa|$. From the second line of Eq.~\eqref{ALP} one can deduce the desired equation 
\be \label{rec1}
\lambda \rho -  |\psi |^2 = -i\bar{\psi} \partial_x \frac{\rho}{\bar{\psi}} + \varkappa \rho^2.
\ee
The solution to this equation is sought as a formal series $\rho(x,\lambda) = \sum\limits_{n=1}^\infty\lambda^{-n}\rho_n(x)$,
which provides the following recurrence relation
\be 
\label{conserved_bright}
\rho_{n+1} = - i\bar{\psi}\partial_x  \left(\frac{\rho_n}{\bar{\psi}} \right) + \varkappa  \sum\limits_{k=1}^{n-1} \rho_k \rho_{n-k},
\qquad \rho_1 = |\psi|^2.
\ee
The second (linearly independent) solution of Eq.~\eqref{ALP} can be obtained by the replacements
\be 
\lambda \to -\lambda,\qquad \psi \to \bar{\psi},\qquad F_1 \to F_2.
\ee
Accordingly, we can define function $\tilde{\rho}(x,\lambda) = \sum\limits_{n=1}^\infty(-\lambda)^{-n}\tilde{\rho}_n$,
satisfying the complex-conjugate recurrence relation
\be 
\tilde\rho_{n+1} = - i\psi\partial_x  \left(\frac{\tilde\rho_n}{\psi} \right) + \varkappa  \sum\limits_{k=1}^{n-1} \tilde\rho_k \tilde\rho_{n-k}.
\ee
  
Following \cite{Faddeev}, we construct two linearly independent (Jost) solutions, and combine them into matrices $T_{\pm}$
characterized by the following large-$x$ asymptotics,
\be 
T_{\pm}(x\to \pm \infty) = \left(
\begin{array}{cc}
	e^{\lambda x/2i} & 0 \\
	0 & e^{-\lambda x/2i}
\end{array}
\right),
\ee
yielding
\be \label{tt1}
T_{\pm}(x) =  \left(
\begin{array}{cc}
	1 & -\tilde{\rho}\sqrt{|\kappa|}/\psi \\
	-\rho\sqrt{|\kappa|}/\bar{\psi} & 1
\end{array}
\right)\left(
\begin{array}{cc}
	e^{\lambda x/2 i+ i \varkappa \int^x_{\pm\infty}\rho(y)\mathrm{d} y} & 0 \\
	0 & e^{-\lambda x/2 i - i \varkappa \int^x_{\pm\infty}\tilde\rho(y)\mathrm{d} y}
\end{array}
\right).
\ee
The two asymptotic solutions are in turn connected by the transfer matrix $T$, namely
\be \label{tt}
T_- = T_+ T ,\qquad T = \bp
	a(\lambda) & -\bar{b}(\lambda) \\
	b(\lambda) & \bar{a}(\lambda)
\ep ,
\ee
with normalization relation $|a(\lambda)|^2 + |b(\lambda)|^2 = 1$.
From Eqs. \eqref{tt1} and \eqref{tt} we immediately get the following integral representation  for $a(\lambda)$
\be 
\log a(\lambda) = i \varkappa \int\limits^\infty_{-\infty}\rho(y,\lambda)\mathrm{d} y ,
\ee
with $\rho(x,\lambda)$ defined in Eq.~\eqref{rec1},

The time dependence of the transfer matrix is derived in \cite{Faddeev}, and is given by
\be 
\frac{\mathrm{d}T(\lambda)}{\mathrm{d}t} = \frac{i \lambda^2}{2}[\sigma_3,T(\lambda)] \qquad \Longrightarrow \qquad
T(\lambda,t) = e^{i\lambda^2t\sigma_3/2}T(\lambda,0)e^{-i\lambda^2t\sigma_3/2},
\label{Ttime}
\ee 
which, in particular, implies that $a(\lambda)$ is an integral of motion,
\be 
\partial_t a(\lambda) = 0,
\ee
which serves as the generating function for an infinite tower of conserved charges $Q_{k}$,
\be 
\log a(\lambda) = i \varkappa \sum_{n=0}^{\infty}\frac{Q_{k}}{\lambda^{k}},
\qquad Q_{k} = \int\limits_{-\infty}^\infty \mathrm{d} x \rho_n(x) =  0,\qquad \frac{\mathrm{d}}{\mathrm{d} t}Q_{k}=0.
\ee
The explicit form of the local densities $\rho_{n}$ can be found with the help of Eq. \eqref{Ttime}
\be 
\partial_t \rho_n  = -\partial_x \left[
i\frac{\partial_x \bar{\psi}}{\bar{\psi}} \rho_n+\rho_{n+1}
\right]  \equiv - \partial_x j_n .
\ee 
The time dependence \eqref{Ttime} induces the following transformation on the Jost solutions 
\be 
T_\pm(0) \to T_\pm(t) = e^{-i\lambda^2t\sigma_3/2}T_\pm(0).
\ee 
The generic solution of the time dependent \eqref{ALPA} is given by the linear combinations of the Jost solutions,
\be  \label{F1F2}
	\begin{split}
	F_1 (x,t,\lambda) & = c_1 \exp \left[ \frac{\lambda}{2 i} x - \frac{\lambda^2}{2 i} t + f \right] + c_2 \left( - \frac{\sqrt{|\varkappa | } \tilde{\rho}}{\psi} \right) \exp \left[ -\frac{\lambda}{2 i} x + \frac{\lambda^2}{2 i} t + \tilde{f} \right], \\
	F_2 (x,t,\lambda)& = c_1 \left( - \frac{\sqrt{|\varkappa | } \rho}{\psi} \right) \exp \left[ \frac{\lambda}{2 i} x - \frac{\lambda^2}{2 i} t + f \right] + c_2 \exp \left[ - \frac{\lambda}{2 i} x + \frac{\lambda^2}{2 i} t + \tilde{f}  \right] .
	\end{split}
\ee
For the purpose of the Darboux transformations employed in Sec. \eqref{subsec:BdG_bright}, we provide the simplified
versions of these expressions for the asypmtotic solution \eqref{asymptBR}. Using that
\be \label{f}
f \to f_a = \frac{i}{2\pi} \int^{x/2t}_{-\infty} \frac{\log \left( 1- \left| b\left( \mu \right) \right|^2 \right)}{\lambda - \mu } \mathrm{d} \mu,
\ee
\be \label{ft}
\tilde{f}\to \tilde{f}_a= \frac{i}{2\pi} \int^{\infty}_{x/2t} \frac{\log \left( 1- \left| b\left( \mu \right) \right|^2 \right)}{\lambda - \mu }  \mathrm{d} \mu ,
\ee
we finally obtain
\be  \label{F12}
\begin{split}
	\mathcal{F}_1 & = \exp \left[ \frac{\lambda}{2 i} x - \frac{\lambda^2}{2 i} t + f_a \right] + \left( - \frac{\sqrt{|\varkappa | } \tilde{\rho}}{\psi} \right) \exp \left[ -\frac{\lambda}{2 i} x + \frac{\lambda^2}{2 i} t + \tilde{f}_a \right], \\
	\mathcal{F}_2 & = \left( - \frac{\sqrt{|\varkappa | } \rho}{\psi} \right) \exp \left[ \frac{\lambda}{2 i} x - \frac{\lambda^2}{2 i} t + f_a \right] + \exp \left[ - \frac{\lambda}{2 i} x + \frac{\lambda^2}{2 i} t + \tilde{f}_a  \right] .
\end{split}
\ee 
We have additionally put $c_1=c_2$ which yields a space-reflection symmetric solution of Eq.~\eqref{dr}.
Notice also that functions $\mathcal{F}_{1,2}$ in Eq.~\eqref{dr} are evaluated for the purely imaginary $\lambda=\lambda_1$,
so there is no ambiguity in the integrals in Eqs. \eqref{f},\eqref{ft}.

\section{\label{app:rep_app}Classical Inverse Scattering Method for 1D Repulsive NLSE}

In this appendix we provide a formal solution to the auxiliary linear problem for the repulsive case \eqref{ALP2}.
The main difference compare to the attractive case is that now we impose nontrivial boundary conditions \be\label{behav}
\lim_{x\to -\infty}\psi(x)= 1,\qquad
\lim_{x\to \infty}\psi(x) = e^{i\theta_a},
\ee 
meaning that the Jost solutions are now characterized by the following asymptotic behavior~\cite{Faddeev}
\begin{align}
T_{-}(x\to -\infty, \lambda) &= \left(\begin{array}{cc}
	1 & -i/w \\ 
	i/w & 1 
\end{array}\right)\exp\left(\sigma_3\frac{k(\lambda)x}{2i}\right),\\
T_{+}(x\to +\infty, \lambda) &= e^{-i\theta \sigma_3/2}T_{-}(x\to -\infty, \lambda) ,
\end{align}
where we have introduced the following functions of the spectral parameter $w$
\be \label{asdf}
\lambda = \frac{w + \zeta /w}{2},\qquad
k  = \frac{w - \zeta /w}{2} = \sqrt{\lambda^2 - \zeta^2}, \qquad \zeta = 2 \sqrt{\varkappa}.
\ee
The formal solution for $T_-(x)$ can be similarly to Eq.~\eqref{tt1} presented as
\be 
T_-(x, \lambda) = \left(\begin{array}{cc}
	1 & \bar{\mathcal{A}}(x) \\ 
	\mathcal{A}(x) & 1 
\end{array}\right)\left(\begin{array}{cc}
e^{\frac{k(\lambda) x}{2i}+i \varkappa \int\limits_{-\infty}^x \rho(y, \lambda) \mathrm{d} y} & 0 \\ 
0 & e^{-\frac{k(\lambda) x}{2i}-i \varkappa \int\limits_{-\infty}^x \rho^*(y, \lambda) \mathrm{d} y} 
\end{array}\right),
\ee
where functions $\mathcal{A}$ and $\rho$ can be found by substituting $T_-$ to Eq.~\eqref{ALP2}, yielding
\be 
\rho(x,\lambda) = -\frac{2}{w} - \frac{i}{\sqrt{\varkappa}} \bar{\psi}(x)\mathcal{A} (x),
\ee
and
\be \label{rec2}
\frac{1}{2} \left(w- \frac{\zeta^2}{w} \right)\rho  - \varkappa \rho^2 +\frac{2 }{w} \partial_x \bar{\psi} \partial_x \frac{\rho}{\bar{\psi}} + 1 - |\psi|^2 =0 .
\ee
The transfer matrix $T(\lambda)$ interpolates between the two solutions $T_-$ and $T_+$,
\be 
T_-(x, \lambda) = T_+(x, \lambda) T(\lambda),
\ee
and admits the following structure
\be \label{scat}
	T(\lambda) = \bp a(\lambda) & \bar{b}(\lambda) \\ b(\lambda) & \bar{a}(\lambda) \ep ,\qquad |a(\lambda)|^2 - |b(\lambda)|^2 = 1.
\ee
From Eq. \eqref{rec2} and the asymptotic behavior for $\psi$ \eqref{behav}, we conclude that $\rho$ vanishes
in the limit $x\to \pm \infty$. Consequently, the asymptotic behavior of the off-diagonal elements is determined by
$\mathcal{A}(x\to +\infty) =ie^{i\theta_a}/w$, and the conserved part of the scattering data \eqref{scat} has the form
\be 
\log [a(\lambda) e^{-i\theta_a/2}] = i \varkappa \int\limits_{-\infty}^\infty \rho(x,k)\mathrm{d} x,\qquad \partial_ta(\lambda) = 0.
\ee
The asymptotic expansion at $k\to \infty$ \eqref{asdf},
\be
\rho(x,k) = \sum\limits_{n=1}^\infty \frac{\rho_n(x)}{k^n},
\label{expand_rho}
\ee
generates an infinite number of conserved charges with densities $\rho_{n}(x)$.
The first few densities can be readily extracted from Eq.~\eqref{rec2}
\be 
\rho_1(x) = |\psi (x) |^2-1,\qquad
\rho_2(x) = -i \bar{\psi}(x)\partial_x\psi(x),\qquad
\rho_3(x) = -\bar{\psi} (x)\partial_x^2 \psi(x)  + \varkappa (1-|\psi (x)|^2)^2,\qquad {\rm etc.}
\ee
The solitons are given by the zeros of $a(w)$ in the upper-half spectral plane. In the solitonless case,
$\tilde{a}(w)\equiv e^{-i\theta/2}a (\lambda(w)) $ is an analytic function in the upper-half $w$-plane \cite{Faddeev}.
Using the Cauchy-Riemann identities, together with the large-$w$ asymptotics $\lim_{|w|\to \infty}\tilde{a}(w)=1+\mathcal{O}(1/|w|)$, 
one can derive the following representation
\be 
\log[\tilde{a}(w)] = \frac{1}{2\pi i} \int \mathrm{d}z \frac{\log |a(z)|^2}{z-w},
\ee
which, using $|a|^2- |b|^2=1$, yields the following expression for the generating function of the local conserved charges
\be 
Q (w) \equiv \int\limits_{-\infty}^\infty \rho(x) \mathrm{d} x  = \frac{1}{2\pi\varkappa}
\int\limits_{-\infty}^\infty  \frac{\mathrm{d} z}{w-z}\log\left(1+|b(z)|^2\right).
\ee 
The inversion transformation $z\mapsto \zeta^2 /z$ leaves $|b(z)|$ invariant and therefore allows us to transform
the domain of integration from $-\infty < z < \infty$ to $|z|>\zeta$. The contributions from the domains
$z>\zeta$ and $z<-\zeta$ are denoted by $Q_+$ and $Q_-$, respectively
\be 
Q(w) = Q_+(w) + Q_- (w).
\ee
The exact spectral representation for $Q_{+}$ is of the form
\be\label{Qplus} 
Q_{+} (w) = \int\limits_{z>\zeta}
\frac{\mathrm{d} z}{2\pi \varkappa} \log(1+|b(z)|^2)\left(\frac{1}{w-z}+\frac{1}{z^2}\frac{1}{w-\zeta^2 /z}\right),
\ee
while for $Q_-$ one has to integrate the same expression for $z<-\zeta$.

Contrary to the attractive case, there is an additional restriction on the scattering data related to the asymptotic phase difference $\theta_a$, which in the solitonless case reads
\be \label{theta}
\theta_a= \frac{1}{\pi}\int\limits_{|z|>\zeta}\frac{\mathrm{d} z}{z} \log(1+|b(z)|^2).
\ee 

\paragraph*{Asymptotics.} Having obtained the exact phase-space expressions for the conserved charges from Eq.~\eqref{rec2},
we next evaluate them on the asymptotic profiles to prove Eq.~\eqref{fr}. There we have argued that the asymptotic expansion
is of the form
\be 
\psi = 1 +\delta \psi_1 +\delta \psi_2 + O\left(\frac{\log t}{t^{3/2}}\right),
\ee
where the leading term $\delta \psi_1 = O(1/\sqrt{t})$ is given by
\be 
\delta\psi_1(x,t) = \frac{f(z)}{\sqrt{zt(z+ \zeta^2 /z)}}\sqrt{2\frac{\mathrm{d} z}{\mathrm{d} \xi}} \left(
z e^{i\theta} - \frac{\zeta^2 e^{-i\theta}}{z} \right) ,\qquad \theta = \varphi + \delta\varphi
\ee
with $z=z(x,t)$ defined in Eqs.~\eqref{lambda_c} and \eqref{k_c}. The subleading term $\delta \psi_2=\mathcal{O}(1/t)$ was
added to fix the first conserved charge.  

The asymptotic expansion for the field $\delta \psi$ results in a similar asymptotic expansion
for the density via Eq.~\eqref{rec2}, namely
\be
\rho(x) = R_1+R_2+ O\left(\frac{\log t}{t^{3/2}}\right),
\ee 
where the orders of functions $R_{1,2}$ are $R_1=\mathcal{O}(1/ \sqrt{t})$ and $R_2 = \mathcal{O}(1/t)$, respectively.
More specifically, Eq.~\eqref{rec2} leads to  
\be 
\frac{1}{2} \left( w- \frac{\zeta^2}{w} \right) R_1   + i \partial_x R_1 - \frac{2 i}{w} \partial_x (\delta \bar{\psi}_1) +\delta \psi_1 + \delta \bar{\psi}_1 = 0 ,
\ee
\be 
 \frac{1}{2} \left( w- \frac{\zeta^2}{w} \right) R_2 + i \partial_x R_2 + i \delta \bar{\psi}_1 \partial_x R_1 - i \partial_x (R_1 \delta \bar{\psi}_1) 
	+ \frac{i}{w} \partial_x (\delta \bar{\psi}^2_1 - 2 \delta \bar{\psi}_2) - \varkappa R_1^2 - \delta \psi_1 \delta \bar{\psi}_1 - \delta \psi_2 - \delta \bar{\psi}_2 = 0.
\ee

From the first equation, taking into account that in the leading order $\partial_x \theta = k = (z- \zeta^2 /z)/2$, we conclude that
\be 
R_1 = 4 (z-\zeta^2 /z)\left[\frac{e^{i\theta}}{w-z}+\frac{e^{-i\theta}}{w-  \zeta^2 /z }  \right]\frac{f(z)}{\sqrt{zt(z+\zeta^2 /z)}}\sqrt{2\frac{\mathrm{d} z}{\mathrm{d} \xi}}.
\ee
This form indicates that the contribution for the conserved charges from $R_1$ is vanishingly small, that is
\be 
 \int \mathrm{d}x R_1(x,t) = o(t).
\ee
Indeed, this follows from the observation that function $\theta(x,t)$ does not have any critical points. 
This way, by ignoring all oscillating terms and full derivatives, we obtain
\be 
R_2 = 2 \frac{\varkappa R_1^2 -i \delta\bar{\psi}_1 \partial_x R_1}{w-\zeta^2 /w} + 2 \frac{\delta \psi_1\delta \bar\psi_1+\delta \psi_2+\delta \bar\psi_2}{w-\zeta^2 /w} .
\ee
To get the correct result for $Q_1 = \int \mathrm{d}x (|\psi|^2-1)$ we assume that $\delta \psi_2$ has the form to complete
the square in the last term, meaning that
\be 
2 \frac{\delta \psi_1\delta \bar\psi_1+\delta \psi_2+\delta \bar\psi_2}{w-\zeta^2 /w} = \frac{4}{z t}\frac{\mathrm{d} z}{\mathrm{d} \xi}\frac{f^2(z)(z+\zeta^2 /z)}{w-\zeta^2 /w}.
\ee
By further taking into account that 
\be 
\frac{\varkappa R_1^2 -i \delta\bar{\psi}_1 \partial_x R_1}{w-\zeta^2 /w} 
= \frac{2 f^2(z)}{t}\frac{\mathrm{d} z}{\mathrm{d} \xi}\frac{(z-\zeta^2 /z)^2}{w-\zeta^2 /w}  \frac{w}{(w-z)(w z- \zeta^2)} ,
\ee
and changing variables $t^{-1}\frac{dz}{d\xi} = \frac{dz}{dx}$, we find
\be 
R_2 = 4 f^2(z)\frac{\mathrm{d} z}{\mathrm{d} x}\left( \frac{1}{w-z}+ \frac{1}{w-\zeta^2 /z}\frac{1}{z^2} + \frac{2}{z} \frac{1}{\zeta^2 -w^2} \right).
\ee
Thus the generating function for the conserved charges reads
\be
	Q_+  = \int_{\zeta t}^\infty \mathrm{d} x \rho(x) = 4\int\limits_{z>\zeta}\mathrm{d}z   f^2(z) \left( \frac{1}{w-z}+ \frac{1}{w-\zeta^2 /z}\frac{1}{z^2} + \frac{2}{z} \frac{1}{\zeta^2 -w^2} \right) .
\label{q3}
\ee
This way comparing with the expression \eqref{Qplus}
we see that we can identify 
\be 
f^2(z) = \frac{1}{8 \pi \varkappa} \log(1+|b(z)|^2).
\ee
Notice that the third term in the equation \eqref{q3} vanishes because the linearized profile corresponds to the total phase change $\theta_a =0$ (see Eq. \eqref{theta}).

\section{\label{app:attractive_solu}Solution to BdG equations in Attractive Case}
\label{BdGA}

In this appendix we describe the normal mode decomposition for generic solutions to BdG equation \eqref{linearized_chi},
discuss their normalization and prove their completeness.
One can immediately check, by the direct substitution, that the linear combination 
\begin{align}
\delta \chi (x,t) &= \delta \chi^c(x,t) + \delta \chi^d(x,t), \\
\delta \chi^c(x,t) &= \int_{-\infty}^{\infty} \mathrm{d} k \mathcal{A}^+_k (t) \varphi^+_k(x) + i \int_{-\infty}^{\infty} \mathrm{d} k \mathcal{A}^-_k (t) \varphi^-_k(x),  \\
\delta \chi^d (x,t)  &= c_1(t)\gamma_1(x)+\tilde{c}_2(t)\tilde{\gamma}_2(x)+i(c_2(t)\gamma_2(x)+\tilde{c}_1(t)\tilde{\gamma}_1(x)),
\end{align}
is a solution of \eqref{linearized_chi}.
The eigenmodes for the continuous part $\varphi^{\pm}_k (x)$ are given by
\be 
\varphi^\pm_k(x) = \frac{e^{i k x}}{\sqrt{2\pi}\omega_k} \left[ \big(k+i\nu \tanh (\nu x)\big)^2\pm \nu^2 \sech^2 (\nu x) \right],
\label{eigenmodes_attrac}
\ee
and the time dependence for the expansion coefficients as
\begin{align}
\mathcal{A}^{\pm}_k(t) = \cos(\omega_k t)\mathcal{A}^{\pm}_k(0) \pm \sin(\omega_k t) \mathcal{A}^{\mp}_k (0),
\end{align}
with the dispersion
\be 
\omega_k = k^2 + \nu^2.
\ee
The reflection property,
\be 
\mathcal{A}^{\pm}_k = \bar{\mathcal{A}}^{\pm}_{-k},\qquad \varphi^\pm_k(x) = \bar\varphi^\pm_{-k}(x),
\ee
allows us to present the real and imaginary parts in the form of Eqs.~\eqref{solution_attra} and \eqref{solution_attra1}.
The discrete part of the spectrum and their dynamics is given by
\begin{align}
\tilde{c}_j(t) &= \tilde{c}_j(0),&\qquad c_j(t) &= c_j(t) + 2\nu \tilde{c}_j(0) t,\quad j = 1,2,\\
\label{gamma1}
\gamma_{1} (x) &= \nu \sech (\nu x) \tanh(\nu x),&\qquad \gamma_{2} (x) &= \nu \sech (\nu x), \\
\label{gamma2}
\tilde\gamma_1(x) &= x \gamma_2(x), &\qquad \tilde{\gamma}_2(x) &= \nu^{-1} \gamma_2(x) -x\gamma_1(x).
\end{align}
The discrete modes can be understood as small perturbations of the soliton solution $\psi_s (x,t)$ which no ``energy cost'' in
the BdG equation. More specifically, taking a general one-soliton solution given by Eq.~\eqref{bright_soliton}, any variation
$\partial_n \psi_s$ with respect to parameters $n=\{u,\nu,v,x_0,\varphi_0\}$ satisfies BdG equation \eqref{linearized2}
for $\delta \psi(x,t)$, i.e.
\be 
	i \partial_t (\partial_n \psi_s) = - \partial_x^2 (\partial_n \psi_s) + 4 \varkappa |\psi_s |^2 (\partial_n \psi_s) + 2\varkappa \psi_s^2 (\partial_n \bar{\psi}_s) .
\ee
Therefore, the discrete modes $\delta \chi^d(x,t)$ can be written as linear superposition of discrete modes of $\delta \psi(x,t)$
\be 
	\delta \chi^d \propto e^{-i\nu^2 t} \sum_n c_n  \partial_n \psi_s (x,t) .
	\label{chi_D}
\ee
The discrete modes given by Eqs. \eqref{gamma1},\eqref{gamma2} can be obtained by evaluating the derivatives at suitable
points in the parameter space, $u=\nu$, $v=x_0=\varphi_0 = 0$, yielding
\begin{align} 
	\label{gamma1attra}
	\gamma_{1} (x) &= \nu \sech (\nu x) \tanh(\nu x) \propto e^{-i\nu^2 t} \partial_x \psi_s(x,t) ,\\
	\label{gamma2attra}
	i \gamma_{2}(x) &= i \nu \sech (\nu x)  \propto e^{-i\nu^2 t} \partial_t \psi_s(x,t) ,\\
	\label{gamma3attra}
	\tilde\gamma_2(x) &= \nu^{-1} \gamma_2(x) -x\gamma_1(x),&\quad \tilde\gamma_2(x) + 2 i \nu t \gamma_2 &\propto e^{-i\nu^2 t} \partial_u \psi_s(x,t) , \\
	\label{gamma4attra}
	i \tilde\gamma_1(x) &= i x \gamma_2(x), &\quad i\tilde\gamma_1(x) + 2 \nu t \gamma_1(x) &\propto  e^{-i\nu^2 t} \partial_v \psi_s(x,t)  .
\end{align}

The normalization coefficients in the eigenmodes are chosen to satisfy the orthogonality relations with respect to the scalar product 
\be 
(F,G) = \int_{-\infty}^{\infty} \mathrm{d} x \, \bar{F}(x)G(x).
\ee
For the continuum part, these relations read
\be
(\varphi^+_q, \varphi^-_k)  = (\varphi^-_q, \varphi^+_k) = \delta(k-q),
\label{ortho_C}
\ee
while similarly for the discrete modes,
\be 
(\gamma_1,\tilde{\gamma}_1) = (\gamma_2,\tilde{\gamma}_2) = 1,\qquad (\gamma_1,\gamma_2) = (\tilde{\gamma}_1,\tilde{\gamma}_2) = 0,
\label{ortho_D}
\ee
together with their orthogonality relations
\be
(\gamma_2,\varphi^+_k) = (\tilde{\gamma_2},\varphi^+_k) = 0,\qquad (\gamma_1,\varphi^-_k) = (\gamma_3,\varphi^-_k) = 0.
\label{ortho_mutual}
\ee
This finally permits to prove the completeness of our solutions,
\be 
\int_{-\infty}^{\infty} \mathrm{d} k \, \bar{\varphi}^-_k(x) \varphi^+_k(y) = \delta(x-y) - \gamma_2(x)\tilde{\gamma}_2(y) - \tilde{\gamma}_1(x)\gamma_1(y),
\label{completeness_a}
\ee
and to conclude that we have found the complete spectrum of eigenmodes.

The initial values of the expansion coefficients $\mathcal{A}^{\pm}_k(0)$, $c_{1,2}(0)$ and $\tilde{c}_{1,2(0)}$ are determined from the initial profile $\delta \chi^{(0)}$ with the help of the above orthogonality relations. Namely, one can easily get
\begin{align}
\label{coefficient_continuum}
\mathcal{A}^+_k(0) &= (\varphi_k^-,{\rm Re}[\delta \chi^{(0)}]),& \mathcal{A}^-_k(0) &= (\varphi_k^+,{\rm Im}[\delta \chi^{(0)}]),\\ 
\label{coefficient_disc_attrac1}
c_1(0) &= (\tilde\gamma_1, {\rm Re}[\delta \chi^{(0)}]),& \tilde{c}_1(0) &=  (\gamma_1, {\rm Im}[\delta \chi^{(0)}]),\\
\label{coefficient_disc_attrac2}
\tilde{c}_2(0) &=  (\gamma_2, {\rm Re}[\delta \chi^{(0)}]),& c_2(0) &= (\tilde{\gamma}_2,{\rm Im}[\delta \chi^{(0)}]).
\end{align}
%
Making use of the completeness relation~\eqref{completeness_a}, we give a compact resolution of the time evolved profile,
\begin{align}
{\rm Re}[\delta \chi(x,t)] &= {\rm Re}[\delta \chi^{(0)}(x)]+\int\limits_{-\infty}^\infty \mathrm{d} y\, G^{+-}(x,y;t) {\rm Re}[\delta \chi^{(0)}(y)] +  \int\limits_{-\infty}^\infty \mathrm{d} y\, [2\nu t\, \gamma_1(x)\gamma_1(y)+G^{++}(x,y;t) ]{\rm Im}[\delta \chi^{(0)}(y)],\\
{\rm Im}[\delta \chi(x,t)](x,t) &=  {\rm Im}[\delta \chi^{(0)}(x)]+\int\limits_{-\infty}^\infty \mathrm{d} y\, G^{-+}(x,y;t) {\rm Im}[\delta \chi^{(0)}(y)]  + \int\limits_{-\infty}^\infty \mathrm{d} y\, [2\nu t\, \gamma_2(x)\gamma_2(y)-G^{--}(x,y;t) ]{\rm Re}[\delta \chi^{(0)}(y)],
\end{align}
with the Green's functions
\begin{align}
\label{G12}
G^{\alpha\beta}(x,y;t) &= \int\limits_{-\infty}^\infty \mathrm{d} k (\cos(\omega_k t)-1)\varphi_k^\alpha(x)\bar{\varphi}_k^\beta(y),\\
\label{G11}
G^{\alpha\alpha}(x,y;t) &= \int\limits_{-\infty}^\infty \mathrm{d} k \sin(\omega_k t)\varphi_k^\alpha(x)\bar{\varphi}_k^\alpha(y),
\end{align}
for $\alpha,\beta = \pm$. Moreover, if the initial profile is real (as e.g. in the quenched profile~\eqref{initialAtrractive}),
the time evolution reads
\be 
\delta \chi(x,t) = \delta \chi(x,0)+\int\limits_{-\infty}^\infty \mathrm{d} y\, G^{+-}(x,y;t) \delta \chi(y,0) +
i\int\limits_{-\infty}^\infty \mathrm{d} y\, [2\nu t \gamma_2(x)\gamma_2(y)-G^{--}(x,y;t) ]\delta \chi(y,0).
\ee

\section{\label{app:repulsive_solu}Solution to BdG equations in Repulsive Case}

In this appendix we obtain the normal mode resolution to the BdG equation in the repulsive regime. Similarly to the attractive case,
one can easily check that the following combination of the continuous and discrete modes is a solution of the Eq.\eqref{solL}:
\begin{align}
\hspace{-2mm}{\rm Re}\left[\delta \psi(x,t)\right] &= (c_1 - 2 \zeta c_2 t )\gamma_1(x)+\int\limits_0^\infty \mathrm{d}k \left[\mathcal{A}_k^+(t)\varphi_k^+(x) +\bar{\mathcal{A}}_k^+(t) \phi^+_k(x) \right], \\
\hspace{-2mm}{\rm Im}\left[\delta \psi(x,t)\right] &= c_2\gamma_2(x)+\int\limits_0^\infty \mathrm{d}k \left[\mathcal{A}_k^-(t)\varphi_k^-(x) +\bar{\mathcal{A}}_k^-(t) \phi^-_k(x) \right],
\end{align}
where the time dependence of the expansion coefficients is given by
\be 
\mathcal{A}_k^{\pm} (t) = \cos (\omega_k t) \mathcal{A}^{\pm}_k (0) \pm  \sin (\omega_k t) \mathcal{A}^{\mp}_k (0),
\ee
with
\be
\omega_k = k\sqrt{k^2 +\zeta^2},
\ee
and the explicit expressions for the eigenmodes reading
\be 
\gamma_1(x)=\frac{\zeta}{4} \sech^2 \left( \frac{\zeta x}{2} \right) ,\qquad \gamma_2(x) = 1. 
\ee
and
\begin{align}
\label{varphiplus}
\varphi_k^+(x) &= \frac{e^{ikx}}{2 \sqrt{2\pi k } (k^2+\zeta^2)^{3/4} }[2k^2+2 i \zeta k \tanh(\zeta x/2)+ \zeta^2 \sech^2(\zeta x/2)],\\
\label{varphiminus}
\varphi_k^-(x) &=  \frac{e^{ikx}}{\sqrt{2\pi k} (k^2+\zeta^2 )^{1/4}}[k + i \zeta \tanh(\zeta x/2)] ,
\end{align}
while the other two modes are obtained b complex conjugation $\phi_k^\pm = \bar\varphi_k^\pm$.

One can easily check orthogonality relations between those modes with respect to the scalar product
\be 
(F,G) = \int_{-\infty}^{\infty} \mathrm{d} x \, \bar{F}(x)G(x).
\ee 
Namely, for the continuous part we have 
\be 
(\varphi^+_q, \varphi^-_k)  = (\varphi^-_q, \varphi^+_k) = \delta(k-q),  \qquad (\phi^+_q, \phi^-_k)  = (\phi^-_q, \phi^+_k) = \delta(k-q),
\label{ortho_dark1}
\ee
\be 
(\varphi^+_q, \phi^-_k)  = (\phi^-_q, \varphi^+_k) =(\phi^+_k,\varphi^-_q)  = (\varphi^+_k,\phi^-_q) =0.
\label{ortho_dark2}
\ee
The discrete modes satisfy normalization
\be 
(\gamma_1 , \gamma_2) = 1,
\ee
and are orthogonal to the continuous spectrum
\be 
(\gamma_1 , \varphi^-_k) = (\gamma_1 , \phi^-_{k}) = 0,\qquad (\gamma_2 , \varphi^+_k) = (\gamma_1 , \phi^+_{k}) = 0. 
\ee
These relations allow us to find the initial values of the expansion coefficients $\mathcal{A}^{\pm}_k(0)$, $c_{1,2}$,
from a given initial profile $\delta \psi^{(0)} = \psi(x,0)$, namely
\begin{align}
\label{continuum_const_dark1}
\mathcal{A}^+_k (0) & = (\varphi^-_k, {\rm Re}[\delta \psi^{(0)}], \qquad & \mathcal{A}^-_k (0) & = (\varphi^+_k, {\rm Im}[\delta \psi^{(0)}]) \\
\label{discrete_const_dark}
c_1 & = (\gamma_2,{\rm Re}[\delta \psi^{(0)}] ) , \qquad & c_2 & = (\gamma_1, {\rm Im}[\delta \psi^{(0)}]) .
\end{align} 
Notice that this procedure makes sense only in all the integrals are convergent, requiring initial potentials which decay
fast enough.

The completeness relation should be understood as the following limit 
\be \label{complR}
\lim\limits_{\epsilon\to 0}\left(\int\limits_{0-i\epsilon}^\infty \mathrm{d}k \bar{\varphi}_k(x)\varphi_k(y) +
\int\limits_{0+i\epsilon}^\infty \mathrm{d}k \bar{\phi}_k(x)\phi_k(y)\right) = \delta(x-y) - \gamma_1(x)\gamma_2(y),
\ee
where the $\epsilon$-prescription regularizes the $1/k$ singularity in the principal value manner.
Furthermore, using relations \eqref{continuum_const_dark1} and \eqref{discrete_const_dark} we can
rewrite solution of the Cauchy problem as
\begin{align}
	{\rm Re}[\delta \psi(x,t)] & = {\rm Re}[\delta \psi^{(0)}(x)]  + \int_{-\infty}^{\infty} \mathrm{d} y G^{+-} (x,y;t)  {\rm Re}[\delta \psi^{(0)}(y)] + \int_{-\infty}^{\infty} \mathrm{d} y [G^{++}(x,y;t)  -2\zeta \gamma_1(x)] {\rm Im}[\delta \psi^{(0)}(y)], \\
	{\rm Im}[\delta \psi(x,t)] & = {\rm Im}[\delta \psi^{(0)}(x)] + \int_{-\infty}^{\infty} \mathrm{d} y G^{-+} (x,y;t)  {\rm Im}[\delta \psi^{(0)}(y)] - \int_{-\infty}^{\infty} \mathrm{d} y G^{--}(x,y;t)   {\rm Re}[\delta \psi^{(0)}(y)] , 
\end{align}
where Green's functions are defined as 
\begin{align}
G^{\alpha \beta} (x,y;t) & = \int_{0}^{\infty} \mathrm{d} k [\cos (\omega_k t) - 1] \left(\varphi^{\alpha}_k (x) \bar{\varphi}^{\beta}_k(y) + {\rm c.c}\right)\\
G^{\alpha \alpha} (x,y;t) & = \int_{0}^{\infty} \mathrm{d} k \sin(\omega_k t) \left(\varphi^{\alpha}_k (x) \bar{\varphi}^{\alpha}_k(y) + {\rm c.c} \right),
\end{align}
for $\alpha$, $\beta = \pm$. Note that these expressions have no singularity at $k=0$.
In addition, for the real initial profile, the time evolution of the linearized fluctuation~\eqref{init_dark} can be simplified as
\be 
	\delta \psi(x,t) = \delta \psi (x,0) + \int_{-\infty}^{\infty} \mathrm{d} y \,  G^{+-} (x,y;t)  \delta \psi (y,0) - i\int_{-\infty}^{\infty} \mathrm{d} y \, G^{--} (x,y;t) \delta \psi (y,0) .
	\label{time_evolution_dark}
\ee

\twocolumngrid

\bibliography{classicalbib}
\bibliographystyle{apsrev4-1}
\end{document}